\documentclass[final]{IEEEtran}

\usepackage{amsthm,amssymb,graphicx,multirow,amsmath,color,amsfonts}%,ulem}
\usepackage[update,prepend]{epstopdf}
\usepackage[noadjust]{cite}
\usepackage[utf8]{inputenc} % was using latin1 before
\usepackage{tikz}
\usetikzlibrary{arrows,calc}		% Optional ticks libraries
\usepackage{bbm} % for \mathbbm{1}
\usepackage{pdfpages}
\usepackage{tabulary}
\usepackage{multirow}
\usepackage{comment}
\usepackage[ruled,vlined]{algorithm2e}
\usepackage{etoolbox}
\usepackage{hyperref}
\usepackage{optidef}
\usepackage[font=small,labelfont=bf]{caption}
\usepackage{subcaption}
\makeatletter
\renewcommand{\SetKwInOut}[2]{%
  \sbox\algocf@inoutbox{\KwSty{#2}\algocf@typo:}%
  \expandafter\ifx\csname InOutSizeDefined\endcsname\relax% if first time used
    \newcommand\InOutSizeDefined{}\setlength{\inoutsize}{\wd\algocf@inoutbox}%
    \sbox\algocf@inoutbox{\parbox[t]{\inoutsize}{\KwSty{#2}\algocf@typo:\hfill}~}\setlength{\inoutindent}{\wd\algocf@inoutbox}%
  \else% else keep the larger dimension
    \ifdim\wd\algocf@inoutbox>\inoutsize%
    \setlength{\inoutsize}{\wd\algocf@inoutbox}%
    \sbox\algocf@inoutbox{\parbox[t]{\inoutsize}{\KwSty{#2}\algocf@typo:\hfill}~}\setlength{\inoutindent}{\wd\algocf@inoutbox}%
    \fi%
  \fi% the dimension of the box is now defined.
  \algocf@newcommand{#1}[1]{%
    \ifthenelse{\boolean{algocf@inoutnumbered}}{\relax}{\everypar={\relax}}%
    {\let\\\algocf@newinout\hangindent=\inoutindent\hangafter=1\parbox[t]{\inoutsize}{\KwSty{#2}\algocf@typo:\hfill}~##1\par}%
    \algocf@linesnumbered% reset the numbering of the lines
  }}%
\makeatother

\SetKwInOut{KwInput}{Input}
\SetKwInOut{KwOutput}{Output}
\def\nb0{{\mathbf{0}}}
\def\nb1{{\mathbf{1}}}

\newtheorem{lemma}{Lemma}

\newtheorem{theorem}{Theorem}

\newtheorem{cor}{Corollary}

\newtheorem{remark}{Remark}

\def\xopt{{\bf x}_{\rm opt}}
\def\popt{{\bf P}_{\rm opt}}

\allowdisplaybreaks % Allows breaking of eqnarray over multiple pages (avoids unnecessary blanks in the document before eqnarray)
\IEEEoverridecommandlockouts
\begin{document}
\graphicspath{{./Figures/}}
\title{
A Probabilistic Reformulation Technique for Discrete RIS Optimization in Wireless Systems
}
\author{\thanks{}}
\author{ Anish Pradhan,~\IEEEmembership{Student Member,~IEEE,} and Harpreet S. Dhillon,~\IEEEmembership{Fellow,~IEEE}
\thanks{A. Pradhan and H. S. Dhillon  are with Wireless@VT, Department of ECE, Virginia Tech, Blacksburg, VA, USA (email: \{pradhananish1, hdhillon\}@vt.edu). The support of U.S. National Science Foundation (Grants ECCS-2030215 and CNS-2225511) is gratefully acknowledged. This paper will be presented in part at the IEEE PIMRC 2023, Toronto, Canada \cite{qualc}.

 %\hfill Manuscript updated: \today.
 } % remove the date for conference drafts
 \vspace{-3mm}
}

\maketitle

\begin{abstract}
The use of reconfigurable intelligent surfaces (RIS) can improve wireless communication by modifying the wireless link to create virtual line-of-sight links, bypass blockages, suppress interference, and enhance localization. However, enabling the RIS to modify the wireless channel requires careful optimization of the RIS phase-shifts. Although discrete RIS is more practical given hardware limitations, continuous RIS phase-shift optimization has attracted significantly more attention than discrete RIS optimization, which suffers from issues like quantization error and scalability. To overcome these issues, we develop a comprehensive probabilistic technique to transform discrete optimization problems into optimization problems of continuous domain probability parameters by interpreting the discrete optimization variable as a categorical random vector and computing expectations with respect to those parameters. We rigorously establish that for the unconstrained case, the optimal points of the reformulation and the original problem coincide. For the constrained case, we prove that the transformed problem is a relaxation of the original problem. We apply the proposed technique to two canonical discrete RIS applications: SINR maximization and overhead-aware rate and energy efficiency (EE) maximization. The reformulation enables both stochastic and analytical interpretations of the original problems, as we demonstrate in our RIS applications. The former interpretation yields a stochastic sampling technique, whereas the latter yields an analytical gradient descent (GD) approach that employs closed-form approximations for the expectation. \textcolor{black}{We have explicitly derived the worst-case computational complexities of the proposed algorithms.} The numerical results demonstrate that the proposed technique is applicable to a variety of discrete RIS optimization problems and outperforms other general approaches, such as closest point projection (CPP) and semidefinite relaxation (SDR) methods.
\end{abstract}
\section{Introduction} \label{sec:intro}
An RIS is a large array composed of low-cost reflecting elements, each of which can impart controllable phase-shifts to the incident signal, thereby modifying the propagation channel. However, because of hardware constraints, the phase-shift induced by each reflective element is normally limited to a set of discrete values. When configured appropriately, RISs can create multiple virtual LoS links \cite{SDR2}, improve channel rank \cite{RI}, transform a fast-fading channel to a slow-fading one \cite{F2S}, suppress co-channel interference \cite{IN}, enhance localization performance \cite{local,emenonye2022ris,emenonye2022fundamentals}, etc. However, optimizing the RIS phase-shifts is the first step to reaping these benefits. Despite the fact that RIS optimization has been extensively investigated in the literature, the majority of its attention has been directed toward the scenario of continuous phase-shifts as this scenario allows easier insights and upper-bounds on the performance of a wireless network. As a consequence, the discrete RIS optimization techniques often appear as an afterthought and the existing techniques that deal with discrete RIS optimizations suffer from various issues, such as scalability and arbitrarily bad performance due to quantization error.

Motivated by the scarcity of scalable and reliable discrete RIS optimization techniques \cite{bpb}, stochastic interpretation of semidefinite relaxation technique \cite{luosdr}, and recent efforts to approach binary optimization problems \cite{aattia} with a lens of probability, we develop a comprehensive technique to transform optimization problems of discrete variables into optimization problems of continuous domain probability parameters. We also rigorously prove that in terms of optimal points, the transformed problem is mathematically equivalent in the unconstrained case and a relaxation with respect to the original problem in the constrained case. Moreover, we gain further insights into our reformulation by investigating the simple two-way partitioning problem and report several moment and gradient results for quadratic forms in binary optimization problems. Ultimately, we apply this reformulation in two different canonical discrete RIS optimization problems demonstrating both the stochastic and analytical approaches. The numerical results confirm that the expectation-based algorithms outperform the conventional approaches. Note that, even though the proposed reformulation is inspired by discrete RISs, the scope of the reformulation is more general and could potentially find applications in other domains as well. 
\subsection{Related Work and Motivation}
Although the study and design of discrete RIS phase-shifts have sparked some interest recently, a large portion of the literature relaxes the discrete constraint to a continuous one, solves the approximate problem, and then quantizes the solution to the closest discrete point. This two-fold approximation is shown to provide arbitrarily bad solutions in the worst-case scenario \cite{bpb}. Yet, continuous RIS optimization remains a big part of the discrete RIS optimization literature. In light of this, some of the most used optimization strategies for continuous RIS phase-shits are combinations of a) SDR, b) minorize-maximization (MM) algorithm, c) penalty methods, d) manifold optimization, e) alternating direction method of multipliers (ADMM), and f) treating phase-shifts as optimization variables instead of the complex gains they provide.

In \cite{SDR1, SDR2}, the authors jointly optimized the active beamforming vector and RIS-based passive beamforming vector in multiple-antenna systems employing SDR. The authors of \cite{SDR3} utilized RISs to enhance the physical layer security of a multiuser multiple-input-single-output (MU-MISO) wireless system. In particular, they used a combination of a penalty-based approach, SDR, and successive convex approximation (SCA) to address the unit modulus constraint of RIS phase-shifts. Energy-efficient RIS designs for a MU-MISO wireless system are developed in \cite{MM1}. In this paper, the authors developed two algorithms to maximize energy efficiency of the network. One of them uses gradient descent method whereas the other uses MM algorithm while considering unit modulus constraint and a realistic power consumption model. Similarly, in \cite{MM2}, the authors leveraged the MM algorithm and complex circle manifold (CCM) method to propose two algorithms that maximizes weighted sum-rate of a multicell MIMO network. In an RIS-assisted backscatter system, the RIS is optimized with a combination of SDR and ADMM technique in \cite{ADMM1}. In \cite{ADMM2}, weighted sum-rate is maximized in an RIS-aided cell-free network through ADMM. The authors of \cite{Penalty1} tackled the resource allocation problem in an RIS-assisted wireless network. They developed an algorithm that uses the SCA and penalty method to jointly optimize phase-shifts and on-off status of RISs to maximize energy efficiency under a total power constraint. The authors of \cite{Manifold1} enhanced the physical layer security by optimizing RIS phase-shifts with fractional programming and manifold optimization techniques. Using a similar technique, the authors of \cite{Manifold2} developed an algorithm combining both SDR and manifold optimization techniques to optimize an RIS-aided edge caching system. Recently, the authors of this paper treated the vector of phase-shifts itself as an optimization variable instead of the vector of the complex gains they provide and optimized the RIS phase-shifts with the GD method to maximize the SINR \cite{arxTHz} similar to the methodology used in \cite{lowc}.

In the realm of discrete RIS optimization literature, the optimization tools often used are exhaustive search, CPP from a continuous relaxation, and branch-and-bound (BB) methods. In \cite{ES1}, the authors investigated a practical discrete RIS-aided wideband orthogonal frequency division multiplexing (OFDM) system and optimized the discrete RIS element-wise exhaustive search in an alternating optimization framework. The authors of \cite{BB1,BB2} optimized the RIS phase-shifts in a MISO wireless network using BB methods that scale exponentially with the number of RIS elements. An RIS-aided MIMO system with low-resolution digital-to-analog converters (DACs) is jointly optimized with particle swarm optimization (PSO) algorithm in \cite{PSO}. This algorithm is shown to work with both continuous and discrete RIS phase-shifts. For maximizing the achievable rate in a MIMO system, the authors in \cite{CPP-PGM} relaxed the discrete RIS constraints to continuous ones and used a projected gradient method (PGM) to solve the problem. Similar continuous relaxation and CPP method was conducted in \cite{CPP2}. Specifically, the authors in \cite{CPP2} investigated the spectral efficiency and energy efficiency trade-off in a MU-MIMO setup. In that paper, the discrete RIS phase-shift constraint was relaxed to a continuous one and then solved by MM and accelerated gradient method. The mentioned optimization problems are either not scalable or suffer from arbitrarily bad performance due to the mentioned two-fold approximation. However, there are some recent efforts \cite{SO1,bpb,SO3} that provide scalable optimal discrete RIS beamforming optimization for single-input-single-output systems as the resulting objective function has a low-rank matrix with $d \leq 2$ based on the fixed-rank result of \cite{SO}. These strategies suffer from being too specialized for single-antenna scenarios and do not work in multi-antenna scenarios that are more practical. 

Going beyond the RIS literature, there has been some interesting approaches to binary quadratic optimization problems \cite{luosdr,PDA,PDA2,aattia,varopt}. The authors of \cite{luosdr} show that SDR formulation is actually a stochastic version of the original non-convex quadratic program. Even when the non-convexity originates from binary variables, the stochastic interpretation still works. The authors of \cite{PDA, PDA2} approach binary quadratic programs with a probabilistic data association (PDA) algorithm that treats the optimization variable as a binary random variable and iteratively updates the probabilities with Gaussian noise approximation. This is shown to achieve near-optimal results. Recently, the authors of \cite{aattia} provide a stochastic gradient descent framework for binary optimization problems by similarly treating the optimization variable as a random variable and then taking expectation on it. Inspired by these probabilistic optimization techniques along with the lack of novel generalized discrete RIS optimization techniques, we develop a comprehensive probabilistic technique to transform discrete optimization problems that opens up new avenues to approach these problems in the continuous probability parameter domain. This technique is then used in discrete RIS cases to showcase its general nature and effectiveness. 
\subsection{Contributions}
We approach the general discrete optimization problems with a different perspective of probability. This results in a comprehensive probabilistic reformulation technique with a wide applicability, including to the discrete RIS problems, which was our original inspiration behind this work. Our key contributions in this paper are listed next.

\subsubsection{A comprehensive probabilistic technique for general discrete optimization problems}
We develop a comprehensive probabilistic technique to reformulate general discrete optimization problems (that are not limited to binary programs) into continuous domain problems. In particular, we re-imagine the entries of the optimization variable as independently but not identically distributed (i.n.i.d) categorical random variables and replace the objective function and constraints, if any, with their expectations. We rigorously establish the equivalence between a general unconstrained problem with a unique optimal solution and the reformulated problem in terms of the optimal point. Additionally, when the original problem is constrained, we prove that the primal solution of the transformed problem is bounded between the dual and primal solution of the original problem. We also show that when strong duality holds, the transformed problem has the same optimal objective value as the original problem. Utilizing this technique, random sampling from a non-degenerate probability parameter solution can provide a better solution with the number of samples similar to Gaussian randomization in SDR.

\subsubsection{Derivation of various analytical moments and their gradients associated with the quadratic form and binary random vectors}
As discrete RIS problems deal with binary phase-shifts often, using our reformulation technique naturally gives rise to expectations associated with the quadratic form and the binary random vectors. For example, both the denominator and numerator of the SINR or secrecy rate often contain quadratic forms \cite{lowc}. The quadratic form is a {\em canonical construct} that appears in the wireless literature frequently. For a gradient-based optimization approach, the gradients of these expectations will also be required. For this reason, we also derive the first and second moments of the said quadratic forms along with their gradients. These key intermediate results are later used in one of our algorithms demonstrating their importance.

\subsubsection{GD algorithm for the SINR maximization}
As the first canonical case study, we apply this technique to an SINR maximization problem and propose a stochastic GD and an analytical GD approach to solve the reformulated problem. We derive and use the first and second-order Taylor approximations of the expectation of the SINR in the analytical GD algorithm while an estimator of the gradient is used in the stochastic approach. The expectation-based algorithms are shown to perform better than the conventional practical approaches evaluated.

\subsubsection{Stochastic sampling approach for ternary random vectors for EE and rate maximization}
We also apply this technique to our second case study, an overhead-aware rate and EE maximization problem which leads to expectations associated with a ternary random vector. As deriving the analytical expectation was challenging for this specific case study, we develop a stochastic sampling approach for such a ternary random vector where the gradient is estimated with Monte Carlo (MC) samples, thereby demonstrating the versatility of the proposed approach. We demonstrate that this framework is well-suited for non-smooth objective forms and performs well in both interference-free and interference-rich scenarios. Moreover, the developed stochastic approach is demonstrated to work with different objective functions like rate and EE without the need for changing the algorithm.
\textcolor{black}{
\subsubsection{Computational complexity discussion} We have also derived worst-case computational complexities with big-O notation for all the proposed algorithms.}
 
\subsubsection*{Notations}
The distribution of a standard complex normal random variable is denoted by $\mathcal{CN}(0,1)$. The matrix, scalar and vector entities are denoted by $\bf X$, $x$, and $\bf x$, respectively. All the vectors are column vectors unless defined explicitly. For a vector $\bf x$, ${\rm diag}\left({\bf x}\right)$ denotes a diagonal matrix with the entries of $\bf x$ as its diagonal elements. For a matrix $\bf X$, ${\bf X}^H$, ${\bf X}^T$, ${\rm Re}\left({\bf X}\right)$, ${\rm Tr}\left({\bf X}\right)$, ${\rm diag}({\bf X})$, and ${\bf X}\succeq0$ denote its conjugate transpose, transpose, real part, trace, diagonal elements as a vector, and positive semidefiniteness, respectively. Additionally, ${\bf X}_{wd} ={\bf X}-{\rm diag}({\bf X})$. The expectation operation is denoted by $\rm E[\cdot]$, ${\rm var}(\cdot)$ denotes a total variance operator which evaluates the trace of the variance-covariance matrix of the random vector argument, and the operator $\odot$ denotes element-wise multiplication between two matrices. The L0 and L2 norm are denoted by $\|\cdot\|_0$ and $\|\cdot\|_2$, respectively. The identity matrix and all-one column vector of dimension $N$ are denoted by ${\bf I}_N$ and ${\bf 1}_N$, respectively.

%%%%%%%%%%%%%%%%%%%%%%
\section{Probabilistic Reformulation for Discrete Optimization} \label{sec:SysMod}
%%%%%%%%%%%%%%%%%%%%%%
\subsection{The Case of Unconstrained Discrete Optimization Problem}
We begin with a general unconstrained discrete optimization problem where we make no assumptions about the objective function's convexity. The optimization variable is a vector of length $n$ and each of the entry can take a discrete value among the set $\mathcal{C}=\{c_1,c_2,\hdots,c_b\}$.

\begin{mini}|s|
{{\bf x} \in \mathcal{C}^n}{f({\bf x}).}{}{}\label{eq:ori}
\end{mini}

Our main goal is to reformulate the problem in a form that does not deal with the discrete domain and shares the optimal solution with the original problem. To that end, we propose to re-imagine entries of $\bf x$ as i.n.i.d categorical random variables with the following joint probability density function (PDF):
\begin{align}
\mathbb{P}({\bf x}|{\bf P})\!=\!\prod\limits_{i=1}^n\sum\limits_{j=1}^b\! \delta(x_i\!-\!c_j)p_{i,j}, p_{i,j}\in[0,1],\! \sum\limits_{j=1}^b p_{i,j}=1, \label{eq:JPDF}
\end{align}
where the $(i,j)$-th entry of the matrix $\bf P$ is denoted by $p_{i,j}$, the $i$-th entry of $\bf x$ is denoted by $x_i$, and $\delta(\cdot)$ is the Dirac delta function. We then reformulate the original problem into a stochastic optimization problem:
\begin{mini}|s|
{p_{i,j} \in \mathcal{F}}{\xi({\bf P})={\rm E}_{{\bf x}\sim \mathbb{P}({\bf x}|{\bf P})}\left[f({\bf x})\right],}{}{}
\label{eq:tran}
\end{mini}%p_{i,j}\in[0,1]
where $\mathcal{F}$ is the set of possible $p_{i,j}$'s defined by \eqref{eq:JPDF}.
The connection between \eqref{eq:ori} and \eqref{eq:tran} and their solution sets are summarised in the following lemma.
\begin{lemma}
The solution sets of the problems \eqref{eq:ori} and \eqref{eq:tran} are denoted by $\Omega_{\bf x}$ and $\Omega_{\bf P}$ and,
\begin{align*}
    \Omega_{\bf x} \subseteq \Omega_{\bf P}.
\end{align*}
Moreover if the unique optimal solution of \eqref{eq:ori} is ${\bf x}_{\rm opt}$, then ${\bf P}_{\rm opt}={\rm Degen}({\bf x}_{\rm opt})$ is the unique optimal solution of \eqref{eq:tran}, where the ${\bf P}={\rm Degen}({\bf x})$ operation implies that the $(i,j)$-th entry of $\bf P$ is defined as $p_{i,j}=1$ only when $x_i=c_j$ while all the other entries are zero. \label{lem:unc}
\end{lemma}
\begin{IEEEproof}
We observe that $\Omega_{\bf x}$ has $b^n$ elements and each of them corresponds to one of the possible $b^n$ combinations that ${\bf x}$ can take. In \eqref{eq:tran}, the same objective values can be attained by the corresponding ${\bf P}={\rm Degen}({\bf x})$ which is the parameter matrix of $n$ degenerate categorical distributions. \textcolor{black}{Let us illustrate this with an example. Suppose we have a vector ${\bf x}=[-1, 1, -1]^T$. Each element of ${\bf x}$ can adopt either $c_1=1$ or $c_2=-1$. By referring to the definition of the ${\rm Degen}(\cdot)$ function given in Lemma \ref{lem:unc}, the resulting parameter matrix for this vector is:
\begin{align*}
{\rm Degen}({\bf x})=\bf P=\begin{bmatrix}
0 & 1 \\ 1 & 0 \\ 0 & 1
\end{bmatrix}.\end{align*}
The interpretation of this matrix is as follows: the first element of ${\bf x}$ is $-1$ with a probability of $1$, the second element is $1$ with a probability of $1$, and so forth. The parameter matrix thereby represents the degenerate distributions of the elements of ${\bf x}$ if it was reimagined as a random vector. Importantly, this parameter matrix can be translated back into the original ${\bf x}$ vector. Consequently, for every possible permutation of ${\bf x}$ (as given in equation \eqref{eq:ori}), there exists a corresponding parameter matrix in equation \eqref{eq:tran} that provides the same objective value.
Upon reflecting on this, it becomes evident that the set $\Omega_{\bf x}$ must be included in $\Omega_{\bf P}$. In mathematical terms, this relationship can be represented as $\Omega_{\bf x} \subseteq \Omega_{\bf P}$.}

For any feasible $\bf P$, it can be shown that,
\begin{align}
\begin{aligned}
\min_{\bf x} f({\bf x})\leq \xi({\bf P})\!=\!\sum\limits_{k=1}^{b^n} f({\bf x}\{k\})\mathbb{P}({\bf x}\!=\!{\bf x}\{k\}|{\bf P}) \leq \max_{\bf x} f({\bf x}), \label{eq:ineq}
\end{aligned}
\end{align}
where ${\bf x}\{k\}$ denotes the $k$-th combination out of possible $b^n$ combinations of $\bf x$. This stems from the observations that the expectation is nothing but a convex combination of all the possible values of $f({\bf x})$. \textcolor{black}{This is possible because the probability terms, which are always nonnegative, sum up to one, enabling the expression of the expectation as this sum. Such a convex combination of scalar values essentially represents a probability-weighted average. Each scalar is weighed by its corresponding probability or chance of occurrence. These probabilities fundamentally dictate the placement of the weighted average on the line between the minimum and maximum scalar values. Due to the constraint of the probabilities adding up to one, this average cannot exist outside this range. For instance, when the probabilities tend to favor larger scalar values, the resulting combination leans closer toward the maximum and vice versa. It is crucial to clarify that equating the expectation to a convex combination does not imply that the expectation is a convex function. We are not discussing the convexity of the expectation itself. Yet, due to the inherent properties of convex combinations, inequality \eqref{eq:ineq} consistently holds, as indicated in \cite{convexgeometry}. Additionally, Carathéodory's theorem offers further proof of this fact \cite{aattia,convexgeometry}.}

Now assume that $\xopt$ is the unique optimal solution of $\eqref{eq:ori}$. It follows that, $\popt=\rm{Degen}(\xopt)$ is an optimal solution of \eqref{eq:tran}. Consider that $\exists {\bf P}_0\neq\popt$, such that, $\xi({\bf P}_0)=\xi(\popt)=f(\xopt)$. The parameter matrix $\bf P$ cannot denote $n$ degenerate categorical distributions as the corresponding ${\bf x}_0={\rm Degen}^{-1}({\bf P}_0)$ would violate the uniqueness assumption on $\xopt$. We then consider the non-degenerate distribution case. As the optimal value $p^*$ is shown to be the same for both of these problems, we can assume that $p^*=f({\bf x}\{k_0\})$ without any loss of generality. Then,
\begin{align}
    &\xi({\bf P}_0)\!=\!\!\sum\limits_{k=1}^{b^n}\! f({\bf x}\{k\})\mathbb{P}({\bf x}={\bf x}\{k\}|{\bf P}_0)=f({\bf x}\{k_0\})=p^* \label{eq:ndp} \\
    &\implies \!\!\!\!\!\!\!\sum\limits_{k=1, k\neq k_0}^{b^n}\!\!\!\!\!\!\!\! \left(f({\bf x}\{k\})-f({\bf x}\{k_0\})\right)\mathbb{P}({\bf x}={\bf x}\{k\}|{\bf P}_0)=0.\label{eq:ndp2}
\end{align}
As for some $k$, the value $f({\bf x}\{k\})$ needs to be equal to $f({\bf x}\{k_0\})$ for \eqref{eq:ndp2} to be true, this would also violate the uniqueness assumption on $k_0$.
\end{IEEEproof}
\subsection{The Case of Constrained Discrete Optimization Problem}
Next, we explore whether such a coincident optimal solution through such a reformulation is valid for constrained problems as well. To that end, we write a general optimization problem with constraints without assuming convexity below:
\begin{mini}|s|
{{\bf x} \in \mathcal{C}^n}{f_0({\bf x}),}{}{}\label{eq:cons}
\addConstraint{f_i({\bf x})}{\leq 0 \quad \forall i=1, 2, \hdots, m}
\addConstraint{h_j({\bf x})}{= 0 \quad \forall j=1, 2, \hdots, r,}
\end{mini}
where the optimal value and solution set of this problem are denoted by $p_c^*$, and $\Psi_{\bf x}$, respectively. The transformed formulation is expressed as:
\begin{mini}|s|
{p_{i,j} \in \mathcal{F}}{{\rm E}_{{\bf x}\sim \mathbb{P}({\bf x}|{\bf P})}[f_0({\bf x})],}{}{}\label{eq:eqns}
\addConstraint{{\rm E}_{{\bf x}\sim \mathbb{P}({\bf x}|{\bf P})}[f_i({\bf x})]}{\leq 0 \quad \forall i=1, 2, \hdots, m}
\addConstraint{{\rm E}_{{\bf x}\sim \mathbb{P}({\bf x}|{\bf P})}[h_j({\bf x})]}{= 0 \quad \forall j=1, 2, \hdots, r,}
\end{mini}
with optimal value $p_e^*$, and solution set $\Psi_{\bf P}$.

\begin{lemma}
The original problem \eqref{eq:cons} and the transformed problem $\eqref{eq:eqns}$ share the same dual problem with the dual solution $d^*$. Moreover,
\begin{align}
    d^* \leq p_e^* \leq p_c^*.
\end{align}
\end{lemma}
\begin{IEEEproof}
\textcolor{black}{Similar to the proof of Lemma \ref{lem:unc}, it can be readily seen that for every feasible $\bf x$ in \eqref{eq:cons}, there is a corresponding feasible parameter matrix ${\bf P}={\rm Degen}({\bf x})$ in \eqref{eq:eqns}. It directly follows from this observation with similar reasoning in the previous proof that the solution set $\Psi_{\bf x}$ is a subset of $\Psi_{\bf P}$. Consequently, the transformed optimization problem can be seen as a relaxation of the original constrained problem \eqref{eq:cons} and provides a better optimal value. Thus it can be established that $p_e^* \leq p_c^*$.}

%The objective function of \eqref{eq:eqns} is bounded between the maximum and minimum of $f_0({\bf x})$ through \eqref{eq:ineq}. Along with that, $\Psi_{\bf x} \subseteq \Psi_{\bf P}$ as ${\bf P}=\rm{Degen}({\bf x})$ is always a feasible solution to \eqref{eq:eqns}. It follows that $p_e^* \leq p_c^*$. The transformation \eqref{eq:eqns} can be seen as a relaxation of $\eqref{eq:cons}$. 

\textcolor{black}{Next, we investigate the dual function of the original problem by expressing it as the infimum of the Lagrangian \cite{cvxbook,Convexintro}:}
\begin{align}
    g_c({\boldsymbol \lambda, \bf v})\!=\!&\inf_{{\bf x} \in \mathcal{C}^n} L({\bf x},{\boldsymbol \lambda},{\bf v})\notag \\\!=\!&\inf_{{\bf x} \in \{c_1,c_2,\hdots,c_b\}^n} f_0({\bf x})+\sum\limits_{i=1}^m \lambda_i f_i({\bf x}) + \sum\limits_{j=1}^r v_j h_j({\bf x}),
\end{align}
where $\lambda_i$ is the $i$-th entry of $\boldsymbol{\lambda}$ and $v_j$ is the $j$-th entry of $\bf v$. \textcolor{black}{The vectors discussed here are the dual variables related to our problem. We can think of the dual function as a {\em softened} form of equation \eqref{eq:cons}, which consists of more stringent or {\em hard} constraints \cite{cvxbook}. Crucially, for all non-negative vectors $\boldsymbol{\lambda}$, the dual function serves as a consistent lower bound for the optimal value of the primal problem, denoted $p_c^*$. For a more in-depth treatment of this well-known result, readers can refer to \cite{cvxbook}. Next, using Lemma \ref{lem:unc}, we can reformulate the above dual function into the following expression:}
\begin{align}
    g_c({\boldsymbol \lambda, \bf v})\!=\!&\!\inf_{p_{i,j} \in \mathcal{F}} {\rm E}_{{\bf x}\sim \mathbb{P}({\bf x}|{\bf P})}[f_0({\bf x})]+\!\sum\limits_{i=1}^m \lambda_i {\rm E}_{{\bf x}\sim \mathbb{P}({\bf x}|{\bf P})}[f_i({\bf x})] + \notag\\&\sum\limits_{j=1}^r v_j {\rm E}_{{\bf x}\sim \mathbb{P}({\bf x}|{\bf P})}[h_j({\bf x})], \label{eq:dualc}
\end{align}
where the optimal dual solution after maximizing the concave dual function is denoted by $d^*$. \textcolor{black}{We note that the dual function of \eqref{eq:eqns} is equivalent to \eqref{eq:dualc}. Given the lower bound characteristic of the dual function, it is logical to conclude $d^* \leq p_e^*$. Moreover, strong duality ensures equality. Compiling these inequalities, we deduce $d^* \leq p_e^* \leq p_c^*,$ which consequently proves the Lemma. It further implies that the relaxation \eqref{eq:eqns} is non-trivial and it is bounded by $d^*$, given that the dual solution is bounded.}
\end{IEEEproof}
%\subsection{Benefits of this reformulation}
%\subsubsection{Transforming problem structure}
%Our reformulation can change the problem and domain structure through a stochastic optimization approach or analytical expectations. Through such structural changes, we can harness the solution approaches of different genres of problems, such as stochastic programming, box-constrained quadratic programming, etc. Even for constrained problems with weak duality, this reformulation provably results in a non-trivial relaxation where we can try multiple solution approaches for difficult problems.

%\subsubsection{Change in the domain}
%Through the reformulation, we also transform the discrete domain to a bounded continuous one that is more tractable in terms of existing convex optimization methods. 

%\subsubsection{Multiple sub-optimal sampling}
%SDR techniques generally come with a randomization step where one can generate a rank-one feasible solution close to the optimal. This randomization step is done multiple times to generate multiple feasible solutions and only the best feasible solution is chosen. Such a multiple sampling routine is natural to our reformulation. For an approximate non-degenerate joint PDF as a solution to the reformulated problem, we can similarly sample and choose the best feasible solution.

%\subsubsection{Non-smooth objectives}
%As shown in \cite{aattia}, through such a stochastic reformulation, gradient descent optimization on non-smooth objective functions is possible through Monte Carlo methods to estimate the gradient.
\subsection{Discussion on the Two-way Partitioning Example}

In this subsection, we will focus on the simple two-way partitioning problem to demonstrate the technique. \textcolor{black}{We focus on this foundational example to facilitate a more comprehensive understanding and to draw parallels with other probabilistic methods more effortlessly. While the selected problem covers a wide array of applications, such as binary phase beamforming in an RIS-aided network \cite{bpb}, we delve into two more complex applications in Section \ref{sec:OpProblems}.} We begin with the description of the two-way partitioning problem below:
\begin{maxi}|s|
{{\bf x} \in \{-1,1\}^n}{{\bf x}^T{\bf W}{\bf x},}{}{}\label{eq:twp}
\end{maxi}
where ${\bf W} \in \mathbb{R}^{n\times n}$ is a symmetric matrix. Next, we derive our reformulation of \eqref{eq:twp} based on Lemma \ref{lem:unc} below starting with the following result:
\begin{align}
    &{\rm E}_{{\bf x}\sim \mathbb{P}({\bf x}|{\bf p}_x)}\left[\sum\limits_{j=1}^n\sum\limits_{i=1}^n x_i x_j W_{ij}\right]\notag\\&=\mathop{\sum\limits_{j=1}^n\sum\limits_{i=1}^n}_{i\neq j} (2p_{x,i}-1) (2p_{x,j}-1) W_{ij} + \sum\limits_{j=1}^n W_{ii}, \label{eq:expt}
\end{align}
where $\bf p$ is the vector of parameters with $p_{x,i}=\mathbb{P}[x_i=1]$ denoting the $i$-th entry, $x_i$ denotes the $i$-th entry of $\bf x$, and $W_{ij}$ denotes the $i,j$-th entry of $\bf W$. This result directly follows from the facts that ${\rm E}[x_i]=2p_{x,i}-1$,  ${\rm E}[x_i^2]=1$, and the entries are i.n.i.d.
\begin{remark}
Note that, if all the entries of $\bf y$ are either $+1$ or $-1$, $\bf y=\rm{Degen}^{-1}({\bf p}_x)$. In other words, in that case, $\bf y$ is a feasible $\bf x$ and vice versa. \label{rem:degbin} 
\end{remark}
With this result, the transformed problem is as follows:
\begin{maxi}|s|
{{\bf y} \in [-1,1]^n}{{\bf y}^T{\bf W}_{wd}{\bf y},}{}{}\label{eq:twp2}
\end{maxi}
where ${\bf y}=2{\bf p}_x-{\bf 1}$, and ${\bf W}_{wd}$ is the matrix ${\bf W}$ with its diagonal elements set to zero. Note that, we effectively converted a binary quadratic program (BQP) to a non-convex box-constrained quadratic program (BoxQP) emphasizing the ability of our reformulation to change the structure of a problem while being equivalent in terms of optimal point. However, this is a known result in the optimization community \cite{Rosenberg,reduction} and our reformulation provides a probabilistic proof. Other than this structural change, we can obtain more insights about our reformulation by focusing on the SDR of the original problem which can be derived by considering the following stochastic program by taking expectation on the objective value and the domain of \eqref{eq:twp} \cite{luosdr}:
\begin{maxi}|s|
{{\bf X} \succeq 0}{{\rm E}_{{\boldsymbol \zeta}\sim\mathcal{N}({\bf 0},{\bf X})}\left[{\boldsymbol \zeta}^T{\bf W}{\boldsymbol \zeta}\right],}{}{}
\addConstraint{{\rm E}_{{\boldsymbol \zeta}\sim\mathcal{N}({\bf 0},{\bf X})}\left[{\boldsymbol \zeta}\odot{\boldsymbol \zeta}\right]}{={\bf 1}_n,}
\label{eq:sqcqp}
\end{maxi}
where ${\bf X}$ is an arbitrary symmetric positive semidefinite matrix, and $\boldsymbol \zeta$ is a random vector drawn from a normal distribution with zero mean and covariance $\bf X$. Through the simple observation ${\rm E}_{{\boldsymbol \zeta}\sim\mathcal{N}({\bf 0},{\bf X})}\left[{\boldsymbol \zeta}{\boldsymbol \zeta}^T\right]={\bf X}$, this is equivalent to the classic SDR problem described below:
\begin{maxi}|s|
{{\bf X}}{{\rm Tr}\left({\bf WX}\right)}{}{}
\addConstraint{{\bf X}\succeq 0}{}
\addConstraint{{\bf X}_{ii}=1, \quad i=1,2,\hdots,n.}{} \label{eq:sdr}
\end{maxi}
In the above formulation, the addition of the rank-one constraint ${\rm rank}({\bf X})=1$ would make the problem equivalent to \eqref{eq:twp}. However, this relaxed formulation is solvable in polynomial time unlike \eqref{eq:twp}. SDR can also be seen as a relaxation of the original problem when we allow $x_i$ to be a multidimensional vector with a unit norm. These vectors can be found from The Cholesky decomposition of the solution of SDR. If the angle between two such vectors is really small, that implies which those two entries of $\bf x$ are more likely to be in the same group \cite{goemans}. In other words, the SDR provides us with pairwise correlation information. In contrast, a relaxed version of our reformulation \eqref{eq:twp2} will provide us the probabilities with which each entry of the original vector $\bf x$ will be $+1$ or $-1$. 

The findings of this subsection reveal that our proposed reformulation has the potential to alter the structure of optimization problems. Even if the change is trivial in the case of simple objective forms, it is expected that more complex objective forms will yield non-trivial changes, which can have a significant impact on the efficiency and effectiveness of the optimization process. We will demonstrate these non-trivial structural changes in canonical case studies related to discrete RIS optimization in Section \ref{sec:OpProblems}. Furthermore, the results indicate that although our reformulation differs from SDR in terms of the information it provides, they share similarities in the formulation from a stochastic standpoint. These results encourage further exploration of our approach in more complex objective forms.

%Note that, the reformulation itself is quite trivial to obtain as ${\bf x}^T{\bf W}{\bf x} = {\bf x}^T{\bf W}_{wd}{\bf x} + {\rm Tr}({\bf W})$. With constraint relaxation, we can easily achieve \eqref{eq:twp2}. 

\subsection{Some Useful Results for Quadratic Expressions for Binary Random Vectors}
As most discrete RIS applications deal with binary phase-shift RIS $\{-1,+1\}$ due to its simplicity in operation and implementation, it is only appropriate to derive the analytical moments and their gradients associated with the binary random vectors defined in \eqref{eq:twp2}. They can be used in different optimization contexts with such expectation-based formulations. The higher moment results are motivated by the previous subsection and will be heavily used in the next section. We begin with the covariance matrix next.
\begin{remark}
   For a random vector ${\bf x} \in \{-1,+1\}^n$ with i.n.i.d entries and expectation $ \mathrm{E}[{\bf x}]={\bf y}$, the covariance matrix is
   \begin{align}
       \mathrm{E}[{\bf x}{\bf x}^T]=({\bf y}{\bf y}^T)\odot{\bf E}_m + {\bf I}_N,
   \end{align}
   where ${\bf E}_m$ is the all-one matrix with a hollow diagonal and ${\bf p}$ is defined similarly to \eqref{eq:pdef}.
\end{remark}
Now, we state the first moment and its gradient in Lemma \ref{lem:quadf} without proof due to its trivial nature and partial proof in \eqref{eq:expt}.
\begin{lemma}
For a random vector ${\bf x} \in \{-1,+1\}^n$ with i.n.i.d entries and expectation $ \mathrm{E}[{\bf x}]={\bf y}$, the expectation and the gradient of a sum between a quadratic form and a linear form are
\begin{align}
    \mu_{qf}({\bf G},{\bf z},{\bf y})\!=&\mathrm{E}[{\bf x}^T{\bf G}{\bf x}\!+\!{\bf z}^T{\bf x}]\!=\!{\bf y}^T{\bf G}_{wd}{\bf y} +{\rm Tr}({\bf G}) + {\bf z}^T{\bf y}, \\
    \vartheta_{qf}({\bf G},{\bf z},{\bf y})=& \nabla_{\bf y} \mathrm{E}[{\bf x}^T{\bf G}{\bf x}] = ({\bf G}_{wd}+{\bf G}_{wd}^T){\bf y} + {\bf z}.
\end{align}
where $\bf G$ is a real symmetric matrix. %Include in notations
\label{lem:quadf}
\end{lemma}
Next, we derive an expectation that is very important for covariance calculations between a quadratic form and a linear form in the next theorem.
\begin{theorem}
For a random vector ${\bf x} \in \{-1,+1\}^n$ with i.n.i.d entries and expectation $ \mathrm{E}[{\bf x}]={\bf y}$, the expectation of a product between a quadratic form and a linear form is
\begin{align}
   &\mu_{ql}({\bf G},{\bf z},{\bf y})= \mathrm{E}[{\bf x}^T{\bf G}{\bf x} {\bf z}^T{\bf x}]=2{\bf y}^T{\bf G}_{wd}{\bf z} +{\bf z}^T{\bf y}{\rm Tr}({\bf G})+\notag\\&{\bf 1}^T\{({\bf G}_{wd}{\bf Y}_{wd})\odot{\bf Y}_{wd}\}({\bf y}\odot{\bf z}),
\end{align}
where $\bf G$ is a real symmetric matrix and ${\bf Y}={\bf y}{\bf 1}^T$. %Include in notations
\label{theo:quadlin}
\end{theorem}
\begin{IEEEproof}
See Appendix \ref{sec:quadlinProof}. 
\end{IEEEproof}
We just state the gradient of the above expectation without proof in Corollary \ref{cor:gradquadlin}.
\begin{cor}
    The gradient of the derived expectation in Theorem \ref{theo:quadlin} can be calculated as:
    \begin{align}
        &\vartheta_{ql}({\bf G},{\bf z},{\bf y})=2{\bf G}_{wd}{\bf z}+{\bf z}{\rm Tr}({\bf G})+(({\bf G}_T^T \odot {\bf E}_m){\bf y})\odot{\bf z} +\notag \\
        &  {\rm diag}({\bf G}_T  {\rm diag}({\bf y} \odot {\bf z}) {\bf E}_m) +({\bf G}_T \odot {\bf E}_m)({\bf y} \odot {\bf z}),
    \end{align}
    where ${\bf G}_T = {\bf G}_{wd}{\bf T}_0$, and ${\bf T}_0={\rm diag}({\bf y}){\bf E}_m$. \label{cor:gradquadlin}
\end{cor}
Next, we focus on the second moment of a quadratic form in Theorem \ref{theo:quadsquare}.
\begin{theorem}
For a random vector ${\bf x} \in \{-1,+1\}^n$ with i.n.i.d entries and expectation $ \mathrm{E}[{\bf x}]={\bf y}$, the second moment of a quadratic form is
\begin{align}
    &\mu_{qs}({\bf G},{\bf y})\!=\!\mathrm{E}[({\bf x}^T{\bf G}{\bf x})^2]={\bf y}^T\left({\bf G}_s\!-\!{\bf F}({\bf y})\right){\bf y}\! +\! {\rm Tr}({\bf G})^2 +\notag\\ & 2{\rm Tr}({\bf Z}) +({\bf y}^T{\bf G}{\bf y})^2\! -\! {\bf d}^T {\bf G}_g {\bf d} ,
\end{align}
where $\bf G$ is a real symmetric matrix, ${\bf d}={\bf y}\odot{\bf y}$, ${\bf G}_s=2{\rm Tr}({\bf G}){\bf G}_{wd}+4{\bf Z}_{wd}$, ${\bf Z}={\bf G}_{wd}{\bf G}_{wd}^T$, ${\bf F}({\bf y})=({\bf y}\odot{\bf y})^T {\rm diag}({\bf G})({\bf G}+{\bf G}_{wd})+4{\bf U}_{wd}$, ${\bf U}= [{\bf I}_N \otimes ({\bf y} \odot {\bf y})^T]{\bf B}$, and ${\bf G}_g=2{\bf G}_{wd}\odot{\bf G}_{wd}$. The matrix ${\bf B}$ is defined through blocks as
\begin{align}
  {\bf B}=  \begin{bmatrix} 
 {\bf b}_{1,1}, \ldots, {\bf b}_{1,N} \\
 \cdots, \cdots, \cdots, \\ 
 {\bf b}_{N,1}, \ldots, {\bf b}_{N,N}.
\end{bmatrix}, 
\end{align}
where the $i$-th element of ${\bf b}_{k,j}$ is ${\bf b}_{k,j}^i=G_{{wd}_{ij}}G_{{wd}_{ki}}$. \label{theo:quadsquare}
\end{theorem}
\begin{IEEEproof}
See Appendix \ref{sec:quadsquareProof}.
\end{IEEEproof}
Now, we derive the gradient of the second moment in the Corollary \ref{cor:grads}. 
\begin{cor}
    The gradient of the derived expectation in Theorem \ref{theo:quadsquare} can be calculated as:
    \begin{align}
        &\vartheta_{qs}({\bf G},{\bf y})=({\bf G}_s + {\bf G}_s^T){\bf y} + 2{\bf y}^T{\bf G}{\bf y}({\bf G} + {\bf G}^T){\bf y} -\notag \\
        &
        2{\bf y}^T({\bf G} + {\bf G}_{wd}){\bf y}({\rm diag}({\bf G})\odot{\bf y})\!-\!  {\bf d}^T {\rm diag}({\bf G})({\bf G}\! +\! {\bf G}_{wd}){\bf y}  \notag \\
        & -{\rm diag}({\bf G})^T{\bf d}({\bf G} + {\bf G}_{wd})^T{\bf y} -  2(({\bf G}_g + {\bf G}_{g}^T){\bf d})\odot{\bf y} -\notag \\
        &8{\bf y}\odot{\bf b}_{s} - 4({\bf U}_{wd} + {\bf U}_{wd}^T){\bf y},
    \end{align}
    where ${\bf d}={\bf y}\odot{\bf y}$, and $i$-th entry of ${\bf b}_s$ is ${\bf y}^T {\bf B}_t[i] {\bf y} - {\rm Tr}({\bf B}_t[i])$. The matrix ${\bf B}_t[i]$ can be derived by multiplicating the $i$-th column of ${\bf G}_{wd}$ with the $i$-th row of ${\bf G}_{wd}$. \label{cor:grads}
\end{cor}
\begin{IEEEproof}
See Appendix \ref{sec:gradsProof}.
\end{IEEEproof}

%%%%%%%%%%%%%%%%%%%%%%%%%%%%%%%%%
\section{Applications of The Proposed Reformulation} \label{sec:OpProblems}
%%%%%%%%%%%%%%%%%%%%%%%%%%%%%%%%%
In a MIMO communication scenario, optimizing RIS phase shifts can be a challenging task, particularly when dealing with discrete RISs. Discrete RIS problems are generally more difficult to solve, making it necessary to split the original problem into smaller sub-problems that can be handled separately. Therefore, we focus on the canonical forms of discrete RIS sub-problems that frequently appear in the literature. For a unified treatment, we have chosen a signal model capable of representing a range of RIS-aided scenarios and sub-problems, including a device-to-device communication link, a cellular network where each antenna serves a different user through antenna selection, and a wireless communication link with interferers while the receive beamformer vector remains fixed \cite{arxTHz}. This signal model can be expressed in the following point-to-point representation:
\begin{align}
&y_r=(h_{d_0}+{\bf h}_0^H{\rm diag}(\boldsymbol{\theta}){\bf f}_0)x_{s,0} + \notag\\ &\sum\limits_{i=1}^{N_I}(h_{d_i}+{\bf h}_i^H{\rm diag}(\boldsymbol{\theta}){\bf f}_i)x_{s,i} + w, \label{eq:CommonSignal}
\end{align}
where $y_r$ is the received signal from the Tx of interest (denoted by $i=0$), $h_{d_i}$ denotes the direct channel between the $i$-th Tx and Rx, ${\bf h}_i$ is the Tx-RIS channel, ${\bf f}_i$ denotes the RIS-Rx channel, $x_{s,i}$ is the data for the $i$-th Tx, ${\rm E}[x_{s,i}^2]=\beta_i$, $\boldsymbol{\theta}$ is the $N$-element discrete RIS phase configuration vector, $N_I$ is the number of interferers, and $w$ is the additive noise. \textcolor{black}{Note that the users and interferers always transmit at their maximum power $p$.} For a general MIMO communication scenario, these channels can be seen as the actual channels pre-multiplied and post-multiplied by precoding and receiver beamformer vectors, respectively. With such a versatile signal model, two use cases for RIS-assisted wireless communication systems are explored, namely, SINR maximization and overhead-aware RIS optimization. Note that, in both cases, RISs are assumed to be controlled by the receiver through an RIS controller \cite{RISarch}.

\subsection{SINR Maximization with RIS Optimization}

\subsubsection{System model} 
We consider a generic system model dictated by the signal model \eqref{eq:CommonSignal}. We consider that the RIS phase vector $\boldsymbol{\theta}=[\theta_1 \quad \theta_2 \hdots \theta_n \hdots \theta_N]^T$ and $\theta_n \in \{-1,+1\}$. For ease of notation, we also define ${\bf h}_{c_i}=\left({\bf h}_i^H{\rm diag}({\bf f}_i)\right)^H$  With this discrete RIS, the SINR can be expressed as,
\begin{align}
    \gamma=\frac{\beta_0|h_{d_0}+{\bf h}_{c_0}^H\boldsymbol{\theta}|^2}{\sum\limits_{i=1}^{N_I} \beta_i|h_{d_i}+{\bf h}_{c_i}^H\boldsymbol{\theta}|^2 + \sigma_w^2} = \frac{f_s(\boldsymbol{\theta})}{f_I(\boldsymbol{\theta})}=\frac{\mathbf{\boldsymbol{\theta}}^T{\mathbf{R}_{0}}
\mathbf{\boldsymbol{\theta}}+{\bf c}_0^T{\bf\boldsymbol{\theta}}}{\mathbf{\boldsymbol{\theta}}^T{\bf K} \mathbf{\boldsymbol{\theta}}+{\bf s}^T{\bf\boldsymbol{\theta}}}, \label{eq:SINR}
\end{align}
where ${\bf R}_i = \beta_i{\rm Re}\left({\bf h}_{c_i}{\bf h}_{c_i}^H + \frac{|h_{d_i}|^2}{N} {\bf I}_N\right)$, ${\bf K}=\sum\limits_{i=1}^{N_I}{\bf R}_i+\frac{\sigma_w^2}{N}\mathbf{I}_{N}$, $\sigma_w^2$ is the variance of the additive Gaussian noise, ${\bf c}_i=2\beta_i{\rm Re}({\rm conj}(h_{d_i}{\bf h}_{c_i}))$, and ${\bf s}=\sum\limits_{i=1}^{N_I} {\bf c}_i$. 
\subsubsection{RIS optimization}
\begin{algorithm}
\SetAlgoNoEnd
\SetAlgoLined
\KwInput{$ {\bf R}_i, {\bf c}_i, \varrho, \varepsilon, \epsilon_{th}, \beta_{\rm init}, G~\forall i$}
\KwOutput{$\bar{\mathbf{\boldsymbol{\theta}}}_{{i+1}}$}
Initialize $t=1$, $\delta_{GD}=1$, and $\mathbf{y}_s^{(t)}={\bf y}_{\rm init}$.\\
\While{$\delta_{GD} \leq \epsilon_{th}$\\}
{
Initialize $\beta^{(1)}=\beta_{\rm init}, d_f=-1$.\\
Calculate $\nabla_{{\bf y}_s} \mathcal{J}_l({\bf y}_s^{(t)})$ from \eqref{eq:dJ1} or \eqref{eq:dJ2}.\\

\While{$d_f \leq 0$\\}{${\bf y}_{new}=\mathbf{y}_s^{(t)}-\beta^{(t)} {\nabla}_{\mathbf{y}_s}\left(\mathbf{y}_s^{(t)}\right).$\\
Find $\mathbf{y}_{proj}$ by clipping the vector ${\bf y}_{new}$ in $[-{\bf 1}_N,+{\bf 1}_N]$.\\
$d_f= - \mathcal{J}_l(\mathbf{y}_s^{(t)})-\varepsilon \beta^{(t)}\|\nabla_{{\bf y}_s} \mathcal{J}_l({\bf y}_s^{(t)})\|_2^2+\mathcal{J}_l(\mathbf{y}_{proj}).$\\
$\beta^{(t)}=\varrho\beta^{(t)}$.}
%$\delta_{GD}=\beta^{(t)}\|{\boldsymbol {\nabla}}_{\boldsymbol{\varphi}}\left(\boldsymbol{\varphi}^{(t)}\right)\|_2^2$.\\
$\mathbf{y}_s^{(t+\frac{1}{2})}=\mathbf{y}_s^{(t)}-\beta^{(t)}\nabla_{{\bf y}_s} \mathcal{J}_l({\bf y}_s^{(t)}).$\\
$\mathbf{y}_s^{(t+1)}\in \min\limits_{{\bf v}_y \in [-1,+1]^N} \| {\bf v}_y-\mathbf{y}_s^{(t+\frac{1}{2})}\|.$\\
$t=t+1$.\\
$\delta_{GD}=\|\mathbf{y}_s^{(t+1)}-\mathbf{y}_s^{(t)}\|_2^2.$
}

${\bf p}_s=\frac{\mathbf{y}_s^{(t)}+1}{2}$.\\
Based on this probability parameter vector $\bf p$, sample $G$ RIS phase-shift vectors.\\
Choose the best RIS phase-shift vector $\boldsymbol{\theta}_{best}$ among them based on the resulting SINR.\\
$\bar{\mathbf{\boldsymbol{\theta}}}_{{i+1}}=\boldsymbol{\theta}_{best}$.
\caption{E-GD} \label{algo:Algo3}
\end{algorithm}
In this subsection, our objective is to maximize the SINR given in \eqref{eq:SINR} while the RIS elements are discrete in nature. The optimization problem is described below:
\begin{align}
\min_{\boldsymbol{\theta}\in\{-1,+1\}^N} \quad &  -\frac{f_s(\boldsymbol{\theta})}{f_I(\boldsymbol{\theta})}. \label{eq:GDRIS}
\end{align}
As the domain of this problem is discrete and the problem is a fractional quadratic program, a common way to solve this problem is to relax the discrete domain and then project the solution to the closest discrete point. The relaxed version is solved through GD in \cite{arxTHz}. Note that, we also consider SDR in the simulation results. We approach this problem with our reformulation \eqref{eq:tran} and transform this problem into a continuous domain problem. The reformulated problem is as follows:
\begin{align}
\min_{{\bf y}_s\in[-1,+1]^N} \quad &  -{\rm E}_{\boldsymbol{\theta}\sim\mathbb{P}_B({\boldsymbol{\theta}}|{\bf p}_s)} \left[\frac{f_s(\boldsymbol{\theta})}{f_I(\boldsymbol{\theta})}\right] , \label{eq:GDRIS2}
\end{align}
where ${{\bf y}_s}=2{\bf p}_s-1$ and ${\boldsymbol{\theta}}$ is assumed to be distributed with the joint PDF
\begin{align}
    &\mathbb{P}_B({\boldsymbol{\theta}}|{\bf p}_s)\!=\!\prod\limits_{n=1}^N \left( \delta(\theta_n\!-\!1)p_{s,n}+\delta(\theta_n\!+\!1)(1-p_{s,n})\right), \label{eq:pdef}
\end{align}
where $p_{s,n}\in[0,1]$ is the $n$-th entry of ${\bf p}_s$ and $\theta_n \in \{-1,+1\}$. We propose two approaches to solve \eqref{eq:GDRIS2}: a) stochastic sampling approach (SSA), and b) analytical gradient descent approach. The former approach generally does not require an explicit expression of the gradient whereas the latter does. \textcolor{black}{In SSA, a typical gradient estimator, which is based on the log-derivative trick and Monte Carlo sampling as described in \cite{williams}, is often used in the GD algorithm. Following that, we have formulated an approach, SSA-B, which is a special case of SSA for binary variables using the same log-derivative trick and Monte Carlo sampling. We have omitted the details of SSA-B here, as this variant of SSA for binary variables has already appeared in a different context - the Bayesian optimal design of experiments, in \cite{aattia}. For a detailed understanding of SSA-B, readers can refer to \cite[Algorithm 3.1]{aattia}. Next, we detail the analytical optimization approach for this case study.} We explicitly develop the SSA for a non-binary random vector in the next case study, which is a direct result of the general probabilistic reformulation technique that we rigorously devised in Section \ref{sec:SysMod}.

In the analytical gradient descent approach, calculating the direct expectation of a ratio of correlated random variables is difficult. So, we consider the Taylor series approximations of such an expectation \cite{stuart2010kendall}. Both the first-order approximation $\mathcal{J}_1({\bf y}_s)$ and second-order approximation $\mathcal{J}_2({\bf y}_s)$ are stated below:
\begin{align}
    &\mathcal{J}_1({\bf y}_s)=\frac{{\rm E}[f_s(\boldsymbol{\theta})]}{{\rm E}[f_I(\boldsymbol{\theta})]} = \frac{\mu_{qf}({\bf R}_0,{\bf c}_0,{\bf y}_s)}{\mu_{qf}({\bf K},{\bf s},{\bf y}_s)},\notag\\
    &\mathcal{J}_2({\bf y}_s)\!=\!\!\mathcal{J}_1({\bf y}_s) - \frac{{\rm E}[f_s(\boldsymbol{\theta})f_I(\boldsymbol{\theta})]}{{\rm E}^2[f_I(\boldsymbol{\theta})]} + \frac{{\rm E}[f_I^2(\boldsymbol{\theta})]{\rm E}[f_s(\boldsymbol{\theta})]}{{\rm E}^3[f_I(\boldsymbol{\theta})]}.
\end{align}
The second-order approximation requires two additional expectations that are derived along with their gradients in \eqref{eq:beq}.
\begin{figure*}
\begin{align}
    &c_v({\bf y}_s)\!=\!{\rm E}[f_s(\boldsymbol{\theta})f_I(\boldsymbol{\theta})]\!=\!\frac{ \mu_{qs}({\bf R}_0 + {\bf K},{\bf y}_s)\!-\!\mu_{qs}({\bf R}_0 + {\bf K},{\bf y}_s)}{4}\!+\! \mu_{ql}({\bf R}_0,{\bf s},{\bf y}_s) \!+\! \mu_{ql}({\bf K},{\bf c}_0,{\bf y}_s) \!+\! {\bf c}_0^T \left(({\bf y}_s{\bf y}_s^T)\odot{\bf E}_m \!+\! {\bf I}_N\right) {\bf s}, \notag \\
    &\vartheta_{cv}=\nabla_{{\bf y}_s}c_v({\bf y}_s)= \frac{ \vartheta_{qs}({\bf R}_0 + {\bf K},{\bf y}_s)-\vartheta_{qs}({\bf R}_0 + {\bf K},{\bf y}_s)}{4} + \vartheta_{ql}({\bf R}_0,{\bf s},{\bf y}_s) + \vartheta_{ql}({\bf K},{\bf c}_0,{\bf y}_s) +  {\bf s} \odot ({\bf E}_m({\bf c}_0 \odot {\bf y}_s)) + \notag\\& {\bf c}_0 \odot ({\bf E}_m({\bf s} \odot {\bf y}_s)),\quad
    v({\bf y}_s)={\rm E}[f_I^2(\boldsymbol{\theta})]=\mu_{qs}({\bf K},{\bf y}_s) + {\bf s}^T \left(({\bf y}_s{\bf y}_s^T)\odot{\bf E}_m + {\bf I}_N\right) {\bf s} +2\mu_{ql}({\bf K},{\bf s},{\bf y}_s), \notag \\
    & \vartheta_v=\nabla_{{\bf y}_s} v({\bf y}_s) = \vartheta_{qs}({\bf K},{\bf y}_s) + 2{\bf s} \odot ({\bf E}_m({\bf s} \odot {\bf y}_s)) +2\vartheta_{ql}({\bf K},{\bf s},{\bf y}_s). \label{eq:beq}
\end{align}
\hrule
\vspace{-5 mm}
\end{figure*}
Using the definitions in \eqref{eq:beq}, we can express the gradients of the Taylor series approximations as follows:
\begin{align}
    &\nabla_{{\bf y}_s} \mathcal{J}_1({\bf y}_s)\!=\! \frac{\vartheta_{qf}({\bf R}_0,{\bf c}_0,{\bf y}_s)-\mathcal{J}_1({\bf y}_s)\vartheta_{qf}({\bf K},{\bf s},{\bf y}_s)}{\mu_{qf}({\bf K},{\bf s},{\bf y}_s)},\label{eq:dJ1} \\
    &\nabla_{{\bf y}_s} \mathcal{J}_2({\bf y}_s) = \nabla_{{\bf y}_s} \mathcal{J}_1({\bf y}_s) - \frac{\vartheta_{cv}}{\mu_{qf}^2({\bf K},{\bf s},{\bf y}_s)} + \notag \\& \mu_{qf}({\bf R}_0,{\bf c}_0,{\bf y}_s)\left( \frac{\vartheta_v}{\mu_{qf}^3({\bf K},{\bf s},{\bf y}_s)} - \frac{3v({\bf y}_s)\vartheta_{qf}({\bf K},{\bf s},{\bf y}_s)}{\mu_{qf}^4({\bf K},{\bf s},{\bf y}_s)} \right) + \notag \\& \frac{2c_v({\bf y}_s)\vartheta_{qf}({\bf K},{\bf s},{\bf y}_s)}{\mu_{qf}^3({\bf K},{\bf s},{\bf y}_s)}+ \frac{v({\bf y}_s)\vartheta_{qf}({\bf R}_0,{\bf c}_0,{\bf y}_s)}{\mu_{qf}^3({\bf K},{\bf s},{\bf y}_s)}. \label{eq:dJ2}
\end{align}
Note that, they are stated without proof as they can be derived trivially with the basic chain rule. Armed with these gradients, we can develop simple update rules of a projected GD algorithm next:
\begin{align}
    \mathbf{y}_s^{(t+\frac{1}{2})}=&\mathbf{y}_s^{(t)}-\beta^{(t)}\nabla_{{\bf y}_s} \mathcal{J}_l({\bf y}_s^{(t)}),\label{eq:gstep} \\
    \mathbf{y}_s^{(t+1)}\in& \min_{{\bf v}_y \in [-1,+1]^N} \| {\bf v}_y-\mathbf{y}_s^{(t+\frac{1}{2})}\|_2,\label{eq:pstep}
\end{align}
where $\mathbf{y}_s^{(t)}=2\mathbf{p}_s^{(t)}-{\bf 1}$ is the transformed probability vector at the $t$-th iteration, $\beta^{(t)}$ is the step-size and $\nabla_{\bf y} \mathcal{J}_l({\bf y}_s^{(t)})$ is the gradient of the $l$-th order Taylor approximation of the true expectation where $l\in\{1,2\}$. The steps \eqref{eq:gstep} and the \eqref{eq:pstep} are considered gradient step and projection step, respectively. For our box constraints, the projection turns out to be clipping the vector $\mathbf{y}_s^{(t+\frac{1}{2})}$ to $-1$ and $+1$. We also use Armijo-Goldstein (AG) line search \cite{agline} to find a good step-size while avoiding saddle points due to its diminishing nature \cite{vanish}. Complete details of the GD approach are shown in Algorithm \ref{algo:Algo3}. Note that, according to the Remark \ref{rem:degbin}, a feasible discrete $\boldsymbol{\theta}$ is also a feasible $\bf x$ and corresponds to the degenerate PDF itself that generates $\boldsymbol{\theta}$. So, we find the vector that aligns the phases of the reflected signals with the phase of the direct signal:
\begin{align}
    {{\varphi}}^{\rm init}_n=e^{-j\left(\arg({\bf h}_{c_0})_n-\arg(h_{d_0})\right)},\label{eq:heusol} \quad \forall n=1,2,\hdots,N,
\end{align}
where $({\bf h}_{c_0})_n$ denotes the $n$-th element of ${\bf h}_{c_0}$ and project it to $\{-1,+1\}$ for a feasible ${\bf y}_{\rm init}$. After the projected gradient descent, we sample $G$ feasible solutions and choose the best one. The complete procedure is described in Algorithm \ref{algo:Algo3}. Note that the numerical results associated with the case studies will be discussed in the next section.
\subsection{Overhead-aware Rate and EE maximization in an RIS-aided system}
In order to tackle another canonical setting, we now focus on the RIS sub-problems where configuring the RIS to optimize the chosen performance metric requires finding the optimal number of reflecting elements $N_{\rm opt}$ first. However, this is only possible for simpler objective forms \cite{overhead1}. To address this limitation, we introduce a comprehensive stochastic sampling approach that optimizes more complex objective forms, including rate and EE while taking interference into account, circumventing the explicit calculation of $N_{\rm opt}$.

\subsubsection{System Model} Our system model is solely dictated by the signal model in \eqref{eq:CommonSignal}. Along with that, we include the overhead and power consumption models developed in \cite{overhead1,overhead2}. We also assume that each RIS element has the ability to turn off or $\boldsymbol{\theta}\in\{-1,0,+1\}^N$. Note that, this allows us to avoid explicit derivation of $N_{\rm opt}$. We also assume that the estimated channels are reliable. Now, we define the rate of the system considering interference and channel estimation overhead below:
\begin{align}
    R(\boldsymbol{\theta})&=\left(1-\frac{T_E(\|\boldsymbol{\theta}\|_0)+T_F(\|\boldsymbol{\theta}\|_0)}{T}\right)\times\notag \\  & B\log_2\left(\!1+\frac{\beta_0|h_{d_0}+{\bf h}_0^H{\rm diag}(\boldsymbol{\theta}){\bf f}_0|^2}{\sum\limits_{i=1}^{N_I}\beta_i|h_{d_i}\!+\!{\bf h}_i^H{\rm diag}(\boldsymbol{\theta}){\bf f}_i|^2+\sigma_w^2}\right),
\end{align}
where the bandwidth is denoted by $B$, the noise variance is $\sigma_w^2=BN_0$, $N_0$ is the noise power spectral density, total duration of the time slot is denoted by $T$, $T_E(\|\boldsymbol{\theta}\|_0)$ denotes the time taken to estimate the channels, and $T_F(\|\boldsymbol{\theta}\|_0)$ is the feedback duration of the RIS configuration. The channel estimation time and feedback duration time are dependent on the number of RIS elements $\|\boldsymbol{\theta}\|_0$ and are expressed next,
\begin{align}
    &T_E(\|\boldsymbol{\theta}\|_0) = T_0(\|\boldsymbol{\theta}\|_0+1), \notag \\  &
    T_F(\|\boldsymbol{\theta}\|_0)=\frac{\|\boldsymbol{\theta}\|_0 b_F}{B_F \log \left(1+p_F\left|h_F\right|^2 /\left(N_0 B_F\right)\right)},
\end{align}
where $T_0$ is the duration of each pilot tone, and $b_F=2$ is the number of bits used to represent the states of each RIS element. Additionally, $B_F$, $p_F$, and $h_F$ refer to the communication bandwidth, transmit power, and effective channel, respectively, in the feedback phase. Subsequently, the total power consumption can be expressed as
\begin{align}
    P_{t o t}(\|\boldsymbol{\theta}\|_0)&=P_E(\|\boldsymbol{\theta}\|_0)+\left(1-\frac{T_E(\|\boldsymbol{\theta}\|_0)}{T}\right) \mu p+\notag \\  &\frac{T_F(\|\boldsymbol{\theta}\|_0)}{T} \left(\mu_F p_F\!-\!\mu p\right)+\|\boldsymbol{\theta}\|_0 P_{c, n}+P_{c, 0},
\end{align}
where $P_E(\|\boldsymbol{\theta}\|_0)=\frac{P_0T_E(\|\boldsymbol{\theta}\|_0)}{T}$ is the power consumption in the channel estimation phase, $P_0$ denotes the power of each pilot tone, $p$ is the \textcolor{black}{maximum} transmit power in the data transmission phase, $P_{c, n}$ is the power required to operate each RIS element and $P_{c, 0}$ is the static hardware power for the remaining system components. Additionally, $\frac{1}{\mu}$ and $\frac{1}{\mu_F}$ denote the transmit amplifier efficiency in the data transmission and feedback phase, respectively. Finally, the EE of the system is defined by,
\begin{align}
    EE(\boldsymbol{\theta})=\frac{R(\boldsymbol{\theta})}{P_{tot}(\|\boldsymbol{\theta}\|_0)}.
\end{align}
In this system model, the number of RIS elements is chosen to be $N = \min(N_{max}, N_0)$, where $N_{max}$ is a parameter and $N_0$ is the maximum integer for which $T_E(N_0)+T_F(N_0) < T$. This condition assures that the rate is realistic.
\begin{algorithm}
\SetAlgoNoEnd
\SetAlgoLined
\KwInput{System parameters and channels, ${\bf r}_{\rm init}$, $t_{max}$, $\epsilon_t$, $\beta_s$, $N_{e}$, $b_m$}
\KwOutput{${\bf r}^*$}
Initialize $t=1$, $\delta_{SGD}=1$, and $\mathbf{r}^{(t)}={\bf r}_{\rm init}$.\\
Define ${\bf s} = \begin{cases} {\bf p}, & {\bf r} = {\bf q} \\ {\bf q}, & {\bf r} = {\bf p} \end{cases}$ \\
\While{$\delta_{SGD} \leq \epsilon_{t}$ and $t \leq t_{max}$\\}
{
Calculate $\hat{b}_{\bf r}^*$ from Lemma \ref{lem:bopt}. \\
Calculate $\tilde{\bf g}_{\bf r}$ from \eqref{eq:pgrad}.\\
$\mathbf{r}^{(t+\frac{1}{2})}=\mathbf{r}^{(t)}-\beta_s\tilde{\bf g}_{\bf r}.$\\
$\mathbf{r}^{(t+1)}\in \min\limits_{{\bf v}_r \in [0,1]^N} \| {\bf v}_r-\mathbf{r}^{(t+\frac{1}{2})}\|_2$ such that ${\bf r}\leq{\bf 1}_N-{\bf s}$.\\
$t=t+1$.\\
$\delta_{SGD}=\|\mathbf{r}^{(t+1)}-\mathbf{r}^{(t)}\|_2^2.$
}

$\mathbf{r}^*=\mathbf{r}^{(t+1)}$.
\caption{Stochastic sampling approach to optimize ${\bf r} \in \{{\bf p},{\bf q}\}$ \textcolor{black}{for the sub-problems in Algorithm \ref{algo:AlgoBCDSGD}.}} \label{algo:AlgoSGD}
\end{algorithm}
\begin{algorithm}
\SetAlgoNoEnd
\SetAlgoLined
\KwInput{System parameters and channels, $\epsilon$, $G_s$}
\KwOutput{$\mathbf{\boldsymbol{\theta}}^*$} 
Initialize ${\bf q}^*$ with a random vector, $i=0$, $\gamma_0=0$, $\Delta=\epsilon+1$, and $\bar{\mathbf{\boldsymbol{\theta}}}_{{0}}$ with all zeros.\\
 \While{$\Delta>\epsilon$}{
  Obtain $\mathbf{p}^*$ from Algorithm \ref{algo:AlgoSGD} with fixed ${\bf q}^*$.\\
  Obtain $\mathbf{q}^*$ from Algorithm \ref{algo:AlgoSGD} with fixed ${\bf p}^*$.\\
  Generate $G_s$ samples of ${\boldsymbol{\theta}}$ from the obtained ${\bf p}^*$ and ${\bf q}^*$.\\
  Set $\gamma_{i+1} = \frac{1}{G_s}\sum\limits_{g=1}^{G_s} \mathcal{J}(\boldsymbol{\theta}\{g\})$ and $\mathbf{\bar{\boldsymbol{\theta}}}_{{i+1}} = \boldsymbol{\theta}\{g^*\}$, where $g^*$ is the index of the random sample that provides the best objective value. \\
 \lIf{$\gamma_{i+1}\leq\gamma_i$}{$\bar{\mathbf{\boldsymbol{\theta}}}_{{i+1}}=\bar{\mathbf{\boldsymbol{\theta}}}_{{i}}$
   }{}
   {
  
  }
  Evaluate $\Delta=|\gamma_{i+1}-\gamma_{i}|/\gamma_{i}$.\\
  $i=i+1$.
 }
 $\mathbf{{\boldsymbol{\theta}}}^*=\mathbf{\bar{\boldsymbol{\theta}}}_{{i-1}}$.
 \caption{\textcolor{black}{SSA-T (Based on the BCD framework)}} \label{algo:AlgoBCDSGD}
\end{algorithm}

\subsubsection{RIS optimization}
The general optimization problem can be expressed as,
\begin{mini}|s|
{\boldsymbol{\theta} \in \{-1,0,+1\}^N}{\mathcal{J}(\boldsymbol{\theta}),}{}{}\label{eq:OAOpt}
\end{mini}
where $\mathcal{J}(\boldsymbol{\theta})$ can be $-R(\boldsymbol{\theta})$ or $-EE(\boldsymbol{\theta})$. As the objective function is non-smooth and non-convex due to the presence of the interference term and L0 norm, we resort to the Lemma \ref{lem:unc} and reformulate the problem below:
\begin{mini}|s|
{{\bf p},{\bf q} \in (0,1)^N}{{\rm E}_{{\boldsymbol{\theta}}\sim \mathbb{P}_E( \boldsymbol{\theta}|{\bf p},{\bf q})}\left[\mathcal{J}(\boldsymbol{\theta})\right],}{}{}
\addConstraint{{\bf p}+{\bf q}\leq {\bf 1}_N.}{}
\label{eq:OAOpttransformed}
\end{mini}
We assume that the $n$-th element of $\boldsymbol{\theta}$ or $\theta_n$ is an independent categorical random variable and can take the value $+1$ with probability $q_n$ that denotes the $n$-th entry of $\bf q$ and $-1$ with probability $p_n$ that denotes the $n$-th entry of $\bf p$. The joint PDF for the ternary random vector can be expressed as:
\begin{align}
    \mathbb{P}_E(\boldsymbol{\theta}|{\bf p},{\bf q})=\prod\limits_{n=1}^N p_n ^{\frac{\theta_n(\theta_n-1)}{2}} q_n ^{\frac{\theta_n(\theta_n+1)}{2}} (1-p_n-q_n) ^{1-\theta_n^2}.
\end{align}
\textcolor{black}{The presence of coupled optimization variables in \eqref{eq:OAOpttransformed} complicates the problem. To circumvent this, we implement the block coordinate descent (BCD) framework. This method decouples the problem with two coupled variable sets into two tractable sub-problems, each addressing a single set of variables while considering the other fixed. Not only does this approach simplify the problems, but it is also inspired by the ties between Dykstra's algorithm for projections onto intersections of convex sets and BCD \cite{dykstra}. Ultimately, the SSA is used to resolve the sub-problems emerging from the BCD structure, as illustrated in the previous case study.}

We start by taking the gradient of the objective function in \eqref{eq:OAOpt} assuming $\bf q$ is fixed. Note that, all the following derivations can easily be derived when $\bf p$ is fixed by substituting $\nabla_{\bf p}$ with $\nabla_{\bf q}$ and are not derived explicitly. However, we will provide those results in the appropriate lemmas.
\begin{align}
    {\bf g}_{\bf p}\!=&\nabla_{\bf p}{\rm E}\left[\mathcal{J}(\boldsymbol{\theta})\right]\!\overset{(a)}{=}\!\sum\limits_{k=1}^{3^N} \mathcal{J}(\boldsymbol{\theta}\{k\}) \nabla_{\bf p} \mathbb{P}_E( \boldsymbol{\theta}\{k\}|{\bf p},{\bf q})\! \notag \\ \overset{(b)}{=}\!& \sum\limits_{k=1}^{3^N} \left(\mathcal{J}(\boldsymbol{\theta}\{k\}) \nabla_{\bf p} \log \mathbb{P}_E( \boldsymbol{\theta}\{k\}|{\bf p},{\bf q})\right)\mathbb{P}_E( \boldsymbol{\theta}\{k\}|{\bf p},{\bf q})
   \notag \\ \overset{(c)}{=}&{\rm E}\left[\mathcal{J}(\boldsymbol{\theta}) \nabla_{\bf p} \log \mathbb{P}_E( \boldsymbol{\theta}|{\bf p},{\bf q})\right],
\end{align}
where $(a)$ comes from the definition of expectation, $\boldsymbol{\theta}\{k\}$ denotes the $k$-th possible combination out of the possible $3^N$ in an arbitrary indexing order, $(b)$ comes from the identity $\nabla_{\bf p} \log \mathbb{P}_E( \boldsymbol{\theta}|{\bf p},{\bf q})=\frac{1}{\mathbb{P}_E( \boldsymbol{\theta}|{\bf p},{\bf q})}\nabla_{\bf p} \mathbb{P}_E( \boldsymbol{\theta}|{\bf p},{\bf q})$, and $(c)$ converts the summation into expectation. The MC approximation of this gradient for a stochastic optimization approach is:
\begin{align}
    \hat{\bf g}_{\bf p}=\frac{1}{N_e}\sum\limits_{j=1}^{N_e} \mathcal{J}(\boldsymbol{\theta}\{j\}) \nabla_{\bf p} \log \mathbb{P}_E( \boldsymbol{\theta}\{j\}|{\bf p},{\bf q}), \label{eq:pgrad}
\end{align}
where $N_e$ is the number of samples used. For completeness and to reduce the variance of this estimator without increasing $N_e$ drastically, we introduce a baseline $b_{\bf p}$ in the objective function. Such an estimator has the following form,
\begin{align}
    \tilde{\bf g}_{\bf p}=&\frac{1}{N_e}\sum\limits_{j=1}^{N_e} \left(\mathcal{J}(\boldsymbol{\theta}\{j\})-b_{\bf p} \right)\nabla_{\bf p} \log \mathbb{P}_E( \boldsymbol{\theta}\{j\}|{\bf p},{\bf q})\notag\\=&\hat{\bf g}_{\bf p} - b_{\bf p}{\bf d}_{\bf p}, \label{eq:BaseEst}
\end{align}
where ${\bf d}_{\bf p}= \frac{1}{N_e} \sum\limits_{j=1}^{N_e}\nabla_{\bf p} \log \mathbb{P}_E( \boldsymbol{\theta}\{j\}|{\bf p},{\bf q})$. Note that ${\rm E}[{\bf d}_{\bf p}]=0$ as the following results stands:
\begin{align}
    {\rm E}[\nabla_{\bf p} \log \mathbb{P}_E( \boldsymbol{\theta}|{\bf p},{\bf q})]=&{\rm E}\left[\frac{1}{\mathbb{P}_E( \boldsymbol{\theta}|{\bf p},{\bf q})}\nabla_{\bf p} \mathbb{P}_E( \boldsymbol{\theta}|{\bf p},{\bf q})\right]\notag\\\overset{(a)}{=}&\nabla_{\bf p} \sum\limits_{k=1}^{3^N} \mathbb{P}_E( \boldsymbol{\theta}\{k\}|{\bf p},{\bf q}) = 0,
\end{align}
where $(a)$ comes from writing out the expectation in a summation. Using this result, we can also show that both estimators are also unbiased and ${\rm E}[\tilde{\bf g}_{\bf p}]= {\rm E}[\hat{\bf g}_{\bf p}]={\bf g}_{\bf p}$. In the next lemma, we include the key gradient results for the ternary random vector that is instrumental in the stochastic sampling approach.

\begin{lemma}
    The gradient of the log of joint PDF with respect to $\bf p$ and $\bf q$ are:
\begin{align}
    &\nabla_{\bf p} \log \mathbb{P}_E( \boldsymbol{\theta}\{j\}|{\bf p},{\bf q})\notag \\&= \sum\limits_{n=1}^N \left(\frac{\theta_n\{j\}(\theta_n\{j\}-1)}{2p_n}+\frac{(\theta_n\{j\}^2-1)}{1-p_n-q_n}\right){\bf e}_n ,\\
    &\nabla_{\bf q} \log \mathbb{P}_E( \boldsymbol{\theta}\{j\}|{\bf p},{\bf q})\notag \\&= \sum\limits_{n=1}^N \left(\frac{\theta_n\{j\}(\theta_n\{j\}+1)}{2q_n}+\frac{(\theta_n\{j\}^2-1)}{1-p_n-q_n}\right){\bf e}_n,    
\end{align}
where ${\bf e}_n$ is the $n$-th unit vector of length $N$.
\end{lemma}
\begin{IEEEproof}
We start by substituting the joint PDF:
\begin{align}
    &\nabla_{\bf p} \log \mathbb{P}_E( \boldsymbol{\theta}\{j\}|{\bf p},{\bf q}) \notag \\&=  \sum\limits_{n=1}^N \frac{\theta_n\{j\}(\theta_n\{j\}-1)}{2} \nabla_{\bf p}\log p_n \notag \\&+ (1-\theta_n\{j\}^2) \nabla_{\bf p}\log (1-p_n-q_n) \notag \\
    &= \sum\limits_{n=1}^N \left(\frac{\theta_n\{j\}(\theta_n\{j\}-1)}{2p_n}+\frac{(\theta_n\{j\}^2-1)}{1-p_n-q_n}\right){\bf e}_n.
\end{align}
Similarly, the gradient $\nabla_{\bf q} \log \mathbb{P}_E( \boldsymbol{\theta}\{j\}|{\bf p},{\bf q})$ can be derived with ease.
\end{IEEEproof}
With these important gradients available, we find the optimal baseline for the estimator defined in \eqref{eq:BaseEst} in the next lemma.
\begin{lemma}
    The optimal baselines when with respect to $\bf p$ and $\bf q$ are
    \begin{align}
        &b_{\bf p}^* = \frac{N_e}{\sum\limits_{n=1}^N \frac{1}{p_n} + \frac{1}{1-p_n-q_n}  } {\rm E}[ \hat{\bf g}_{\bf p}^T {\bf d}_{\bf p}],\notag \\& b_{\bf q}^* = \frac{N_e}{\sum\limits_{n=1}^N \frac{1}{q_n} + \frac{1}{1-p_n-q_n}  } {\rm E}[ \hat{\bf g}_{\bf q}^T {\bf d}_{\bf q}], 
    \end{align}
where $\hat{\bf g}_{\bf q}$ and  ${\bf d}_{\bf q}$ can be found by replacing the $\nabla_{\bf q}$ in place of $\nabla_{\bf p}$ in \eqref{eq:pgrad} and \eqref{eq:BaseEst}, respectively. \label{lem:bopt}
\end{lemma}
\begin{IEEEproof}
See Appendix \ref{sec:boptProof}.
\end{IEEEproof}
\begin{remark}
    We can also approximate ${\rm E}[ \hat{\bf g}_{\bf p}^T {\bf d}_{\bf p}]$ by taking $b_m$ batches of $N_e$ data points and average them to get $\hat{b}_{\bf p}^*$ to use in the algorithm.
\end{remark}
Now we have all the information to develop the stochastic sampling approach for ternary random variables. \textcolor{black}{The algorithm to solve the sub-problems is demonstrated in Algorithm \ref{algo:AlgoSGD} and the BCD architecture is illustrated in Algorithm \ref{algo:AlgoBCDSGD}.} In the Algorithm \ref{algo:AlgoSGD}, the entries of ${\bf r}_{\rm init}$ are independent and identically distributed (i.i.d) with uniform distribution $\mathcal{U}(0,r_{max})$, where $0< r_{max}\leq1$. By choosing a small $r_{max}$, we control the initial sparsity of the solution.
\textcolor{black}{
\subsection{Worst-case computational complexity discussion}
In this subsection, we derive the worst-case computational complexities for the proposed algorithms in terms of big-O notation. However, we would like to note that the complexity of gradient descent-based algorithms cannot be trivially expressed in the big-O notation, as the number of iterations for convergence heavily depends on the initial point and cannot be precisely determined \cite{antennatilt}. In the literature, the number of iterations is regarded as a parameter, and subsequently, the complexity is represented using big-O notation \cite{lagd,pgdO}. In this subsection, we follow the same approach, while also preserving more terms in the big-O expression for a better comparison among the proposed algorithms. Building upon the previous discussion, The algorithms are based on five fundamental operations: gradient calculation, descent-projection, inner looping, outer looping, and sampling. The descent-projection operation has a complexity of $O(N)$ across all algorithms. Regarding the inner looping operation, we need $I_{E_1}, I_{E_2}, I_1,$ and $I_2$ iterations for the gradient descent algorithms to converge for E-GD using first and second-order Taylor approximations, and for the stochastic sampling approaches with binary and ternary variables respectively. There is typically no need for outer looping iterations except for the ternary variable stochastic sampling method due to the BCD framework. In this case, we assume that $I_{BCD}$ iterations are needed to achieve convergence.
\subsubsection{E-GD with first-order Taylor series approximation}
For this algorithm, the gradient calculation step is primarily dictated by the matrix multiplications inherent in the quadratic forms of \eqref{eq:dJ1}, with a complexity of $O(N^2)$. The sampling step adds an $O(GN^2)$ complexity due to $G$ evaluations of the objective function, making the total complexity $O(I_{E_1}(N^2+N)+GN^2)$.
\subsubsection{E-GD with second-order Taylor series approximation}
For this variant of the algorithm, the gradient calculation step is primarily affected by the matrix multiplications required for computing the matrix ${\bf U}$ as per Theorem \ref{theo:quadsquare}, and has a complexity of $O(N^4)$. The other steps share the same complexities as the first-order version, yielding a total complexity of $O(I_{E_2}(N^4+N)+GN^2)$.
\subsubsection{Stochastic sampling for binary variables}
The first step of this algorithm, the gradient estimator calculation, is dominated by the $N_{ens}$ objective evaluations resulting in a complexity of $O(N_{ens} N^2)$. The sampling step carries a complexity of $O(G_sN^2)$ due to $G_s$ objective function evaluations, which makes the overall complexity $O(I_1(N_{ens}N^2+N)+G_sN^2)$.
\subsubsection{Stochastic sampling for ternary variables}
The main differences between this algorithm and the binary variant lie in the objective function evaluation, which has a complexity of $O(N^2+N)$ due to the additional L0 norm calculation. This yields a total complexity of $O(I_{BCD}I_2(N_{ens}(N^2+N)+N)+G_sN^2)$.
}
\section{Simulation Results}
\textcolor{black}{For our simulation results, we focus on a canonical (and perhaps most practically relevant) RIS scenario where RIS can significantly enhance performance: the creation of virtual line-of-sight (LoS) links when direct paths are obstructed, as highlighted in \cite{MM1,arxTHz,gcthz}. We maintain this assumption throughout our simulation.} We also consider that we operate in a high interference regime where one interferer exists with average power similar to our user. This also highlights the ability of our developed algorithms to cope with high interference. The common simulation parameters used in both the cases are $\beta_i = p\delta_{PL}$, $p$ is the transmit power, $\delta_{PL}=-110$ dB, $B=5$ MHz, and $N_0=-174$ dBm/Hz \cite{overhead1}. Additionally, all the channels are Rician distributed with the Rician factor of $4$ while all the results in this section are averaged over $1000$ independent channel realizations.
\subsection{SINR maximization with RIS optimization}
In this application, we compare the achievable capacity $C_{cap}=\log_2(1+\gamma)$ of our developed algorithms with the popular SDR method and the CPP methods. The transmit power $p$ is considered to be $0$ dBm. The algorithm parameters are $\varrho=0.5,$, $\varepsilon=0.0005$, $\epsilon_{th}=10^{-2}$, $\beta_{\rm init}=0.01$, and $G=100$. Our proposed first-order and second-order analytical GD algorithms are denoted by E-GD-1 and E-GD-2, respectively while our proposed stochastic sampling approach is denoted by SSA-B. The solution of the GD algorithm developed in \cite{arxTHz} for continuous phase shifts projected to the discrete phase-shifts also acts like a baseline and is denoted by CPP-1. The CPP of the solution of \eqref{eq:GDRIS} when the constraint is relaxed to be continuous is denoted by CPP-2. Note that, the only difference between E-GD-1 and CPP-2 is the final sampling step as the former treats the solution as a probability vector, and the latter projects it to $\{-1,+1\}$ for a solution. The CPP of the simple signal alignment scheme in \eqref{eq:heusol} is denoted by SA. CPP methods are considered comparison baselines as they are more practical in terms of speed and are often used in the literature over the traditional branch-and-bound methods that do not scale well with the number of elements.

\begin{figure*}
    \centering
    \begin{subfigure}[t]{0.32\textwidth}
        \centering        \includegraphics[width=\textwidth]{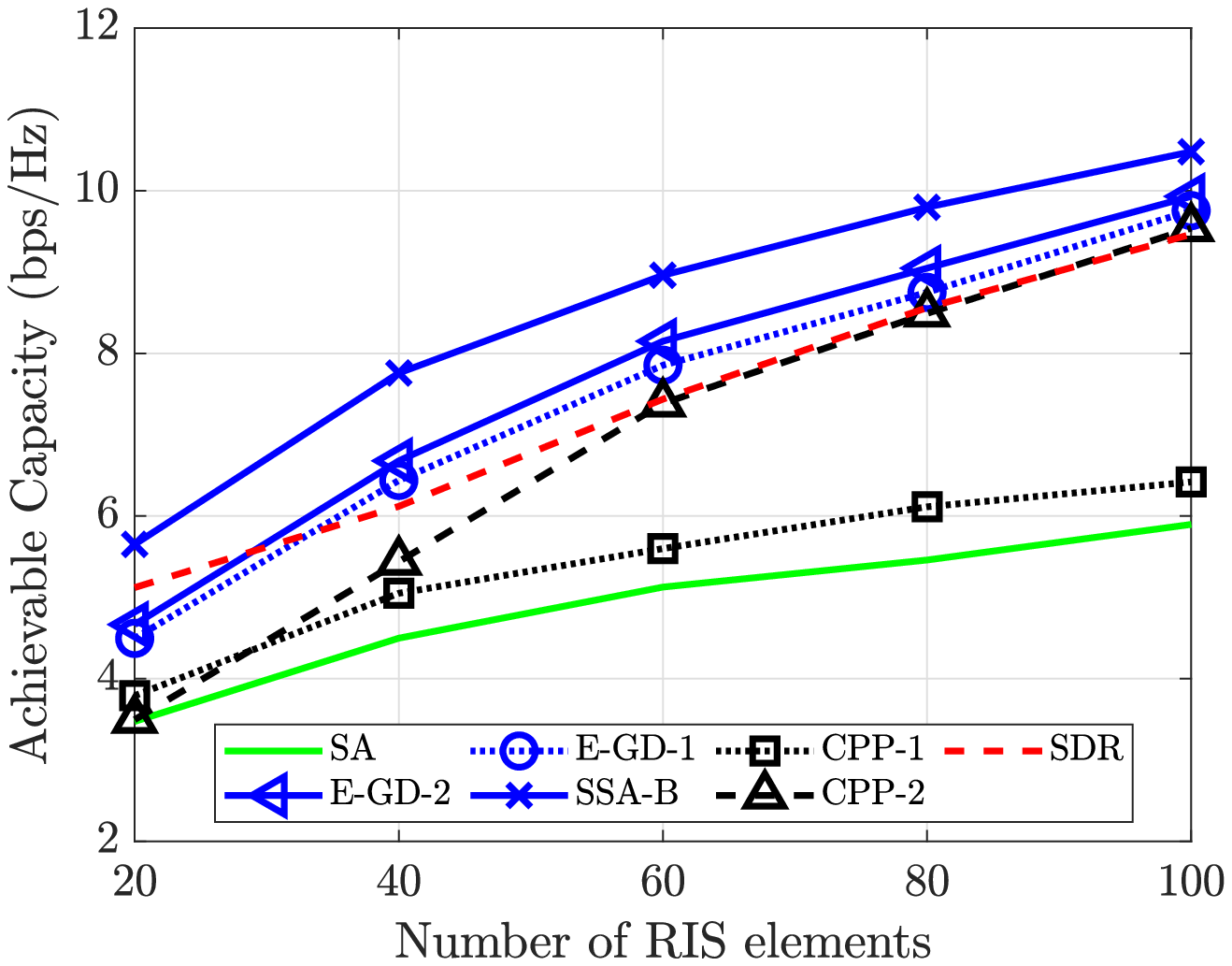}
        \caption{Achievable capacity.}
        \label{fig:Cap}
    \end{subfigure}%
    \begin{subfigure}[t]{0.32\textwidth}
        \centering        
        \includegraphics[width=\textwidth]{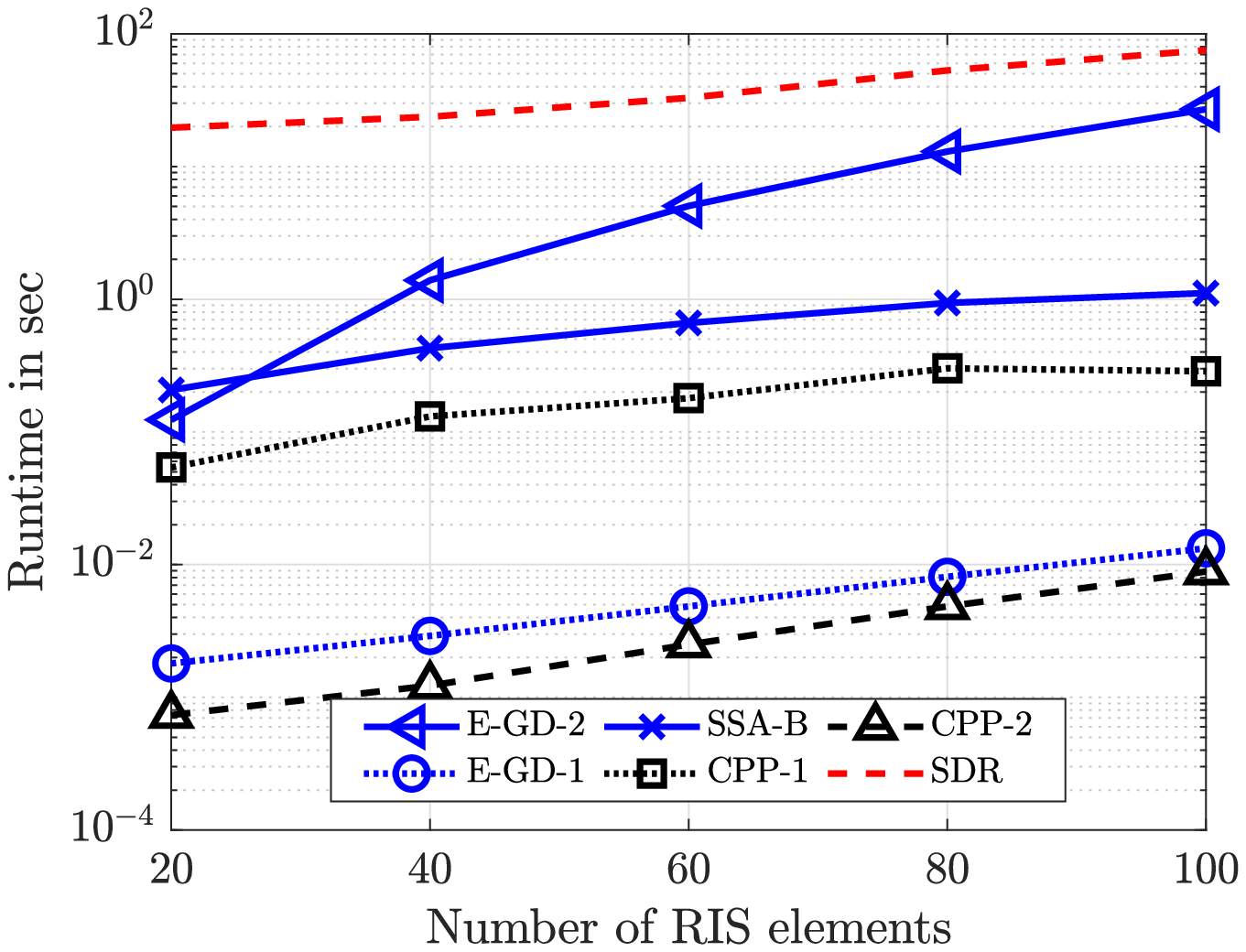}
        \caption{Runtime of algorithms.}
        \label{fig:RT}
    \end{subfigure}
        \caption{Comparison of the algorithms developed.}    
\end{figure*}

In Fig. \ref{fig:Cap}, we can observe that all the expectation-based algorithms perform better than the CPP algorithms, for all $N$, and the SDR for $N>20$. \textcolor{black}{Along with that, SSA-B outshines the expectation-based EGD algorithms that utilize approximations for expectation computation. The edge of SSA-B lies in its robust gradient estimates, derived without reliance on Taylor series approximations, hence providing more precise results. Moreover, the accuracy of SSA-B's gradient estimates can be enhanced by increasing the sample size, although this incurs a higher computational cost.}  \textcolor{black}{Moreover, the scheme CPP-1 performs worse compared to CPP-2 due to its design for continuous RIS phase-shifts with unit-modulus constraints, which means its RIS optimization variable domain spans all angles from $0$ to $2\pi$ corresponding to the set of all unit-modulus complex gains. In contrast, the domain of CPP-2 ranges from $-1$ to $1$, making it closer to the original domain of $\{-1,+1\}$. This difference gets more prominent as the number of RIS elements grows and CPP-2 provides a sharper increase in achievable capacity than CPP-1.}

in Fig. \ref{fig:RT}, we plot the run-time for a single iteration of all the algorithms with varying numbers of RIS elements. These results are taken from the simulations needed to create Fig. \ref{fig:Cap} on a 3.6GHz Intel Core i7-4790 8-CPU system with 16GB RAM. From this plot, we note that the runtime of SSA-B is between the E-GD-1 and E-GD-2 methods while SDR is prohibitively slow. The runtime of our proposed E-GD-2 method is better than SDR but still slower than its first-order counterpart due to the complex gradient calculation. The overall performance of our analytical GD algorithms is dependent on the trade-off between the complexity of the gradient and the accuracy of the approximation for the expectation. These simulation results demonstrate the superiority of the expectation-based algorithms in discrete optimization problems providing important insights into such analytical expectation derivation.

\subsection{Overhead-aware rate and EE maximization in an RIS-aided system}
In this application, we maximize the rate and the EE of the system with our stochastic sampling approach. As a baseline, we compare it with the solution in \cite{overhead1} without interference projected to the discrete RIS phase-shifts. It should be noted that when interference exists, this baseline is no longer relevant because the unimodality required to compute the optimal number of RIS elements is dependent on the simple objective structure without interference. The simulation parameters are set according to \cite{overhead1}: $B_F=1$ MHz, $P_{c,0}=45$ dBm, $P_{c,n}=10$ dBm, $\mu=\mu_F=1$, $T_0=1$ ms, $p_F=30$ dBm, $P_0=10$ dBm, $T=100$ ms, and $N_{max}=300$. The optimization algorithm parameters are, $\epsilon=10^{-6}$, $N_{e}=200$, $b_m=10$, $r_{max}=0.1$, $t_{max}=300$, $G_s=10000$, $\epsilon_t=10^{-8}$, $\beta_s=0.5$ for EE and $\beta_s=0.01$ for rate optimization. In the simulation figures, the upper bound is calculated with the optimum continuous phase shifts without interference through the unimodal approach (UA) devised in \cite{overhead1}. The CPP of this approach also acts as a baseline and is denoted by UA while our algorithm is denoted by SSA-T or stochastic sampling approach for the ternary variable. 

\begin{figure*}
    \centering
    \begin{subfigure}[t]{0.32\textwidth}
        \centering        \includegraphics[width=\textwidth]{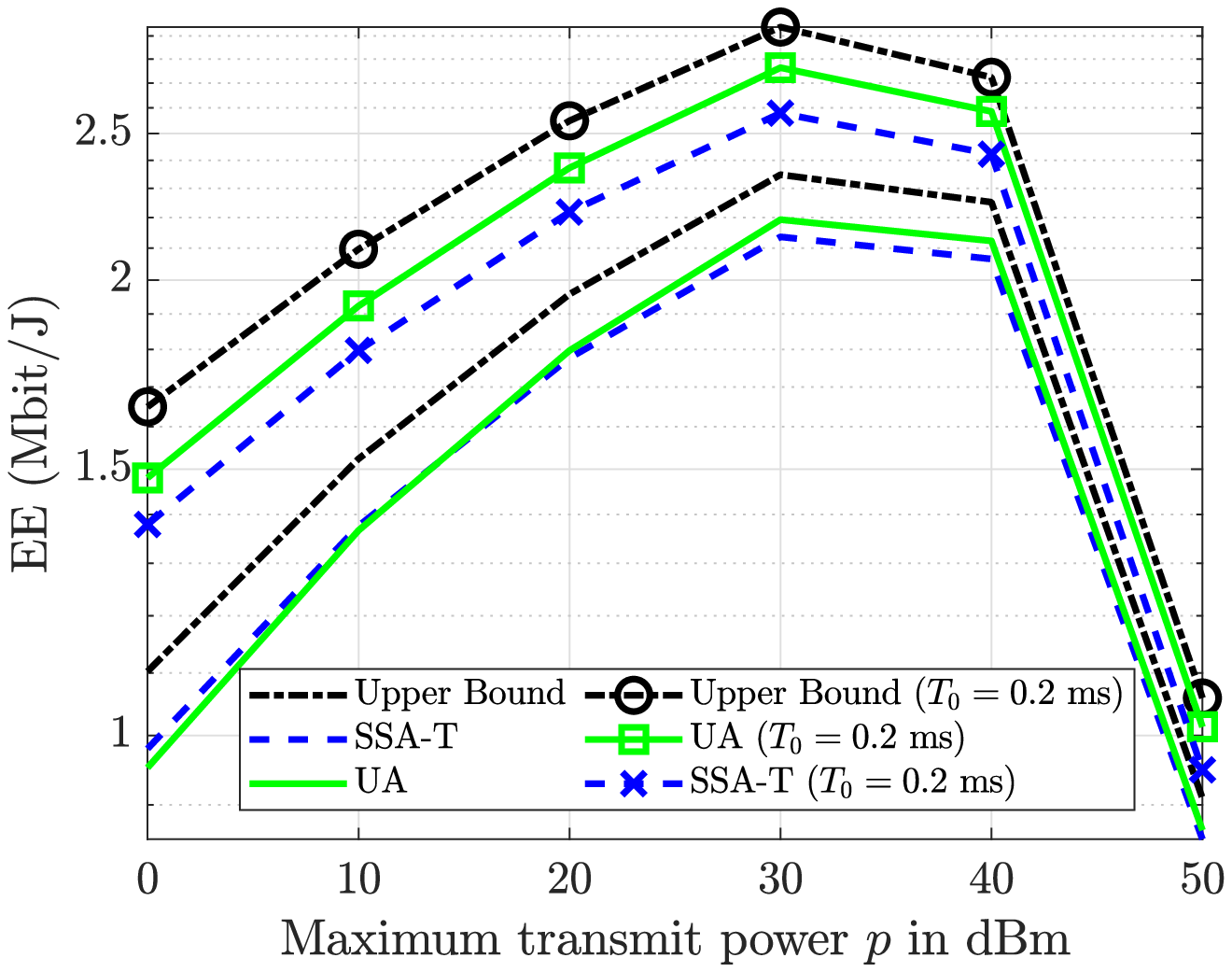}
        \caption{EE vs $p$ without interference.}
        \label{fig:EEmaxWO}
    \end{subfigure}%
    \begin{subfigure}[t]{0.32\textwidth}
        \centering        
        \includegraphics[width=\textwidth]{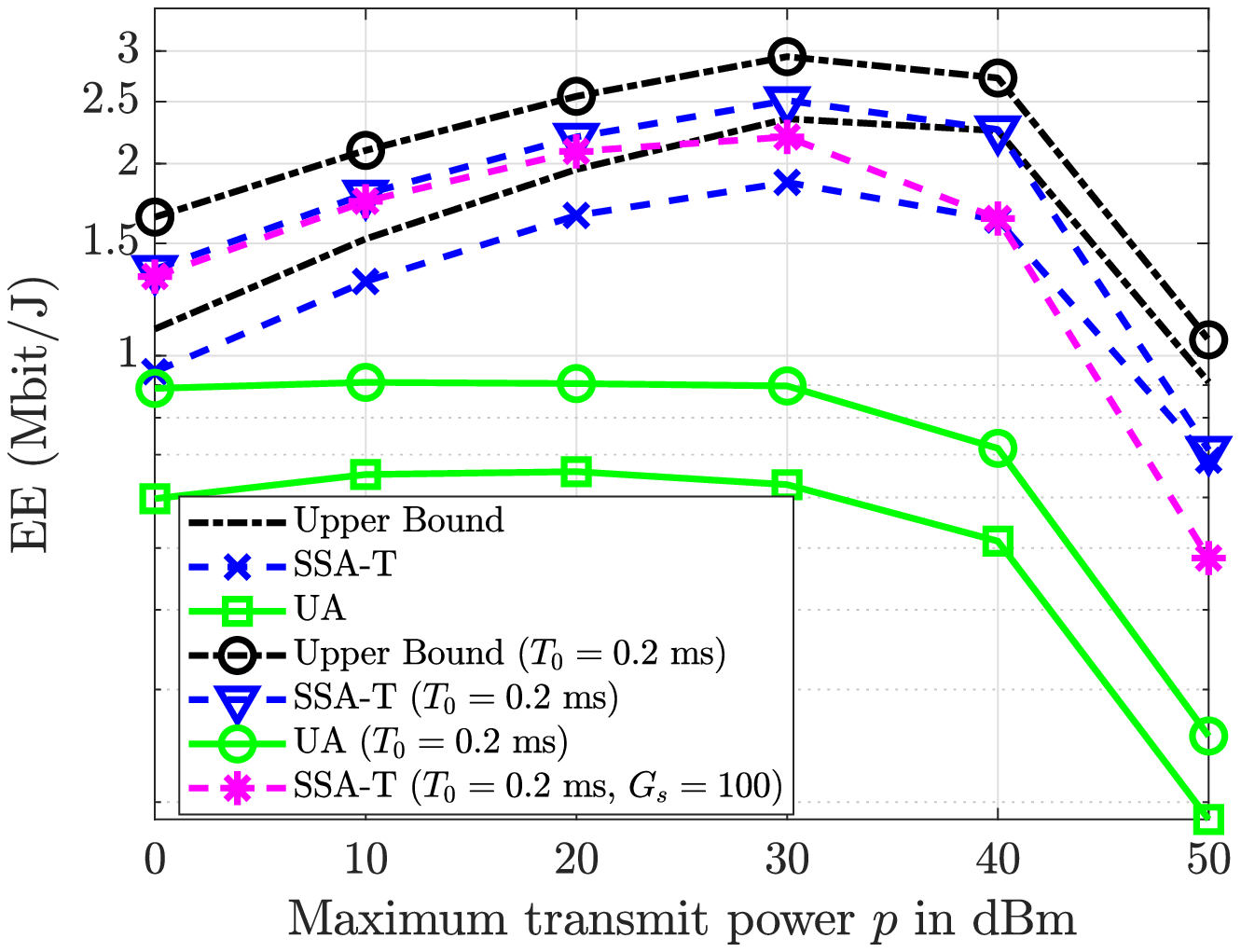}
        \caption{EE vs $p$ with interference.}
        \label{fig:EEmaxW}
    \end{subfigure}
     \begin{subfigure}[t]{0.32\textwidth}
        \centering        
        \includegraphics[width=\textwidth]{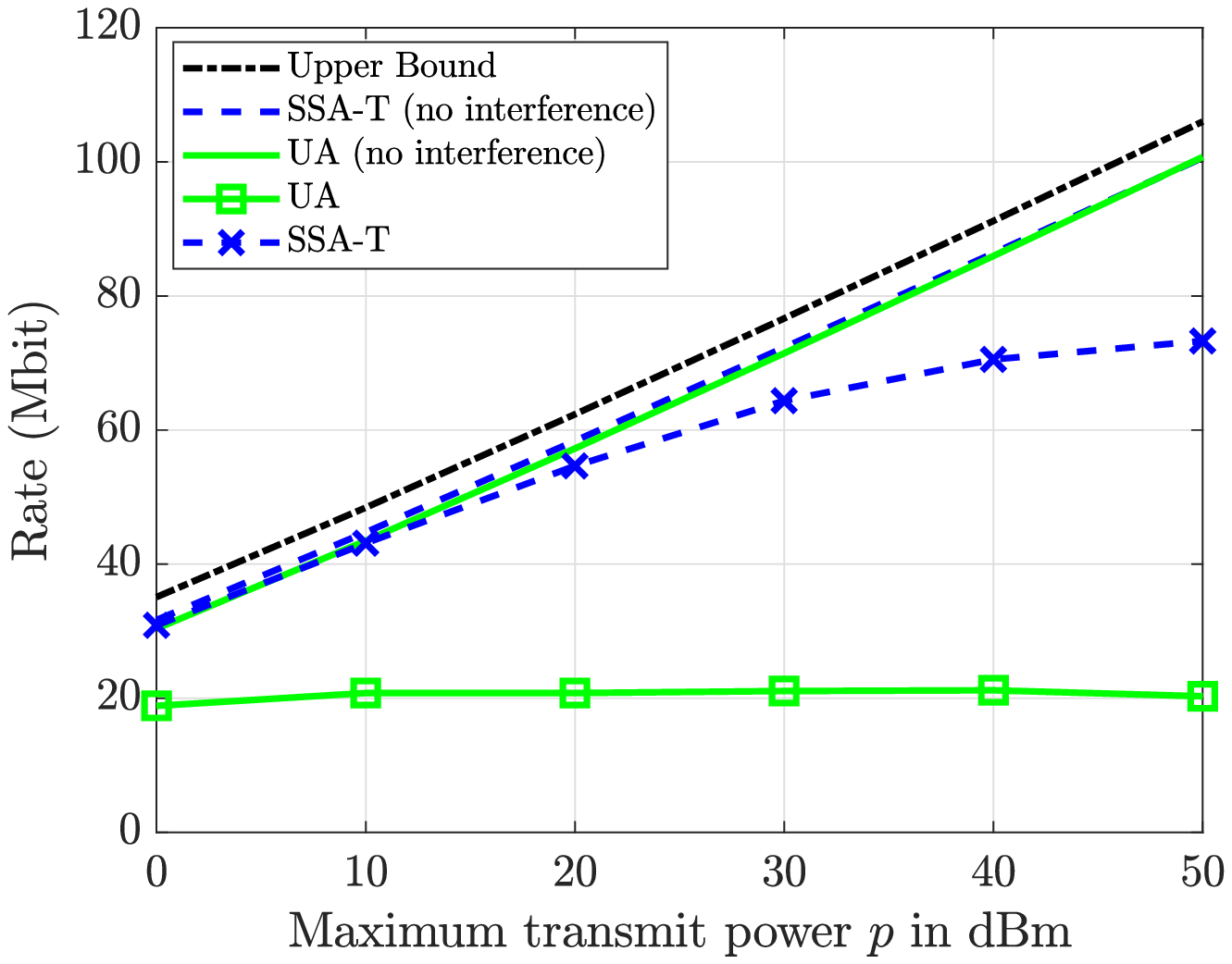}
        \caption{Rate vs $p$.}
        \label{fig:Rmax}
    \end{subfigure}
               \caption{Performance with transmit power $p$ in dBm.}   
\end{figure*}

In Fig. \ref{fig:EEmaxWO}, we plot the average EE achieved with the transmit power when interference is not present. For $T_0=1$ ms, our algorithm performs very similarly to the unimodal approach. However, for $T_0=0.2$ ms, our algorithm achieves an EE that is $0.18$ Mbit/J less at $p=30$ dBm than the UA. \textcolor{black}{While the UA method offers optimal results in the continuous RIS case where no interference is present - a scenario that can be viewed as a special instance of the general formulation with zero interference - it naturally extends well to the discrete case as well. In contrast, our proposed algorithm has a broader scope, demonstrating its capability to handle any form of objective function. Despite this versatility, the trade-off is a guarantee of optimality, hence the observed performance is completely expected. Our algorithm uniquely excels in managing interference and can adapt to any general objective form, an area where the UA method notably underperforms. Therefore, the superior performance of the UA approach in this specific case is anticipated, as it was designed precisely for such interference-free scenarios. This distinction underscores the unique use case of the stochastic sampling approach: when a reliable solution for the continuous problem exists, discrete projection may be sufficient. However, when the objective function becomes complex, even in its continuous form, our proposed algorithm shines, providing high-quality solutions where other methods might fall short.} We can also observe that at the high transmit power region, the EE drops as power consumption dominates and our algorithm approaches the unimodal approach and the upper bound.

\begin{table*}
    \centering
    \begin{tabular}{|c|c|c|c|c|c|c|}
\hline
     & 0 dBm & 10 dBm & 20 dBm & 30 dBm & 40 dBm & 50 dBm\\
     \hline
    UA ($T_0=0.2$ms) & $70.8$ & $59.7$ & $51.7$ & $46.4$ & $49.2$ & $94.5$  \\
    \hline
     UA ($T_0=1$ms) & $21.1$ & $17.3$ & $14.64$ & $13$ & $13.4$ & $23.6$\\
     \hline
    SSA-T ($T_0=0.2$ms) & $52.7$ & $47.4$ & $46$ & $46.1$ & $47.9$ & $40.7$  \\
    \hline
     SSA-T ($T_0=1$ms) & $17.4$ & $14.6$ & $13.5$ & $13.4$ & $13.2$ & $14.7$ \\
     \hline
\end{tabular}
    \caption{Average number of RIS elements to maximize EE in presence of interference with varying $p$.}
    \label{tab:EE}
\end{table*}

In Fig. \ref{fig:EEmaxW}, we plot the average EE achieved with the transmit power when interference is present. \textcolor{black}{In situations where interference is present, as noted in Section \ref{sec:OpProblems}, the transmit power displayed on the x-axis represents the maximum transmit powers of both the user and the interferer. This mimics a non-cooperative scenario where each entity maximizes its own performance by transmitting at peak power simultaneously.} We observe that our algorithms continue to perform close to the upper bound whereas the unimodal approach fails as expected. We can also observe the general trend of better performance with decreasing $T_0$. The reasoning is two-fold: a) we have more time to transmit data due to lower channel estimation time, and b) more overhead for RIS elements can be supported, resulting in the utilization of more RIS elements. This can be verified in Table \ref{tab:EE} where we report the average number of RIS elements to maximize EE in presence of interference varying with the transmit power. We can also observe the effect of $G_s$ in this figure. With a lower $G_s=100$, the EE achieved is around $0.62$ Mbit/J less than the default parameter $G_s=10000$ case at $p=30$ dBm. Finally, in Fig. \ref{fig:Rmax}, we observe that in the rate maximization problem, \textcolor{black}{the proposed algorithm performs closer to the upper bound than the UA approach, irrespective of interference. Without interference, the performance of SSA-T marginally exceeds that of UA. However, in the presence of interference, the advantage of SSA-T over UA becomes substantial. Furthermore, as we increase the maximum transmit power, it becomes clear that the achievable rate via our proposed approach reaches a saturation limit. This limit is imposed by the proportional increase in interference power which is not completely suppressed by the discrete RIS along with user transmit power.}
\section{Conclusion}
In this paper, we developed a novel probabilistic reformulation technique for general discrete optimization problems. In particular, we interpret the discrete optimization variable as a categorical random vector and take expectations on the objective function along with any constraints present. We provide rigorous mathematical justification that the corresponding degenerate PDF of the unique optimal solution of an unconstrained problem is the unique optimal solution of the transformed problem and for a constrained problem, the primal solution of the transformed problem is bounded between the dual and primal solutions of the original problem implying that it is a relaxation of the original problem. However, if strong duality holds, the transformed problem provides the same objective value as the original constrained problem. We also explored a simple two-way partitioning problem to gain more insights into our reformulation such as its similarity to SDR, and capability to change the problem structure. We ultimately used this technique to tackle two canonical discrete RIS applications: a) SINR maximization, and b) overhead-aware rate and EE maximization. As demonstrated in our RIS applications, the reformulation allows for both stochastic and analytical interpretations of the original problems. For the SINR maximization problem, an analytical GD technique based on closed-form approximations for the expectation is proposed, while a stochastic sampling approach is proposed for both applications. The numerical results reveal that there is a fundamental trade-off between the complexity of the gradient and the accuracy of the approximation in our proposed analytical GD methods, and the expectation-based algorithms outperform the other algorithms evaluated. The simulation results also demonstrate that our proposed framework is very general and performs well for both rate and EE maximization problems without much change in the algorithm. In particular, we show that it performs at par with the algorithm specifically developed for the interference-free case when interference is not present and keeps performing well even when interference exists. \textcolor{black}{We also explicitly calculate the worst-case computational complexities for our proposed algorithms.} \textcolor{black}{As the scope of this technique is very general, utilizing this technique to develop a more sophisticated projected gradient descent framework and a general methodology to deal with constrained problems are left as future work.}

%%%%%
\appendix
%%%%%
\subsection{Proof of Theorem \ref{theo:quadlin}} \label{sec:quadlinProof}
   The expectation can be calculated by converting the matrix expressions into series sums as shown below,
    \begin{align}
        &\mathrm{E}[{\bf x}^T{\bf G}{\bf x} {\bf z}^T{\bf x}]= \quad 2\sum\limits_{i=1,i\neq j}^n\sum\limits_{j=1,k=j}^n \mathrm{E}[x_i]{ G}_{ij}{z}_j +  \notag\\ & \notag\mathrm{E}\!\!\left[\sum\limits_{k=1}^n {z}_k x_k \!\!\!\!\sum\limits_{i=1,i=j}^n \!\!\!\!\! x_i^2{G}_{ii}\right]\!+\! \mathrm{E}\!\!\left[\sum\limits_{i\neq j \neq k, k=1}^n  \sum\limits_{j=1}^n \sum\limits_{i=1}^n x_ix_jx_k{G}_{ij}{z}_k\right]\\ &\overset{(a)}{=}
        2{\bf y}^T{\bf G}_{wd}{\bf z} +{\bf z}^T{\bf y}{\rm Tr}({\bf G})+\!\!\!\!\!\!\sum\limits_{i\neq j \neq k, k=1}^n  \sum\limits_{j=1}^n \sum\limits_{i=1}^n y_iy_jy_k{G}_{ij}{ z}_k.
    \end{align}
    In step $(a)$, we use the fact that $\bf G$ is real symmetric, $x_i^2=1$, and ${\rm E}[x_i]=y_i$. The third term can be expressed in the following form:
    \begin{align}
    \sum\limits_{i\neq j \neq k, k=1}^n  \sum\limits_{j=1}^n \sum\limits_{i=1}^n y_iy_jy_k{G}_{ij}{z}_k \!=\!   \begin{bmatrix}
        y_1z_1 \\ y_2z_2 \\ \vdots \\ y_n z_n
    \end{bmatrix}^T \!\! \begin{bmatrix}
        {\bf y}^T{\bf G}_1{\bf y} \\  {\bf y}^T{\bf G}_2{\bf y} \\ \vdots \\  {\bf y}^T{\bf G}_n{\bf y}
    \end{bmatrix},
    \end{align}
    where ${\bf G}_i$ denotes the matrix ${\bf G}$ with the $i$-th row and column set to zeros. In matrix form, this can be expressed as ${\bf 1}^T\{({\bf G}_{wd}{\bf Y}_{wd})\odot{\bf Y}_{wd}\}({\bf y}\odot{\bf z})$, where ${\bf Y}={\bf y}{\bf 1}^T$. This completes the proof.
\subsection{Proof of Theorem \ref{theo:quadsquare}} \label{sec:quadsquareProof}
    We start this proof by expanding a generic quadratic term in \eqref{eq:secondm}.
    \begin{figure*}
    \begin{align}
        &q_s({\bf x})=({\bf x}^T{\bf G}{\bf x})^2=\sum\limits_{l=1}^N\sum\limits_{k=1}^N\sum\limits_{j=1}^N\sum\limits_{i=1}^N x_i x_j x_k x_l G_{ij} G_{kl} \notag \\=& \!\!\!\!\!\!\sum\limits_{k=1, k=l}^N\sum\limits_{j=1}^N\sum\limits_{i=1}^N x_i x_j x_k^2 G_{ij} G_{kk} \!+ \!\!\!\!\!\!\sum\limits_{l=1, k\neq l}^N\sum\limits_{k=1}^N\sum\limits_{i=1,j=i}^N\!\!\!\!x_i^2  x_k x_l G_{ii} G_{kl} + 
        2 \sum\limits_{l=1, k\neq l}^N\sum\limits_{k=1}^N x_k^2 x_l^2 G_{lk} G_{kl} \quad \notag\\&+ 4\sum\limits_{k=1,i \neq j \neq k}^N\sum\limits_{j=1}^N\sum\limits_{i=1,i=l}^Nx_i^2  x_j x_k G_{ij} G_{ki} + \sum\limits_{l=1,i \neq j \neq k\neq l}^N\sum\limits_{k=1}^N\sum\limits_{j=1}^N\sum\limits_{i=1}^N x_i x_j x_k x_l G_{ij} G_{kl}\notag \\=& (\mathbf{x} \odot \mathbf{x})^T {\rm diag} ({\bf G}) \mathbf{x}^T {\bf G}\mathbf{x} + (\mathbf{x} \odot \mathbf{x})^T {\rm diag} ({\bf G}) \mathbf{x}^T {\bf G}_{wd}\mathbf{x} + 2({\bf x}\odot{\bf x})^T {\bf G}_{wd}\odot{\bf G}_{wd} ({\bf x}\odot{\bf x})\quad+
        4\mathbf{x}^T {\bf U}_{wd}\mathbf{x} + \notag \\& \sum\limits_{l=1,i \neq j \neq k\neq l}^N\sum\limits_{k=1}^N\sum\limits_{j=1}^N\sum\limits_{i=1}^N x_i x_j x_k x_l G_{ij} G_{kl}, \label{eq:secondm}
    \end{align}
    \hrule
    \end{figure*}
In \eqref{eq:secondm}, the matrix ${\bf U}= [{\bf I}_N \otimes ({\bf y} \odot {\bf y})^T]{\bf B}$, and the matrix ${\bf B}$ is defined through blocks as
\begin{align}
  {\bf B}=  \begin{bmatrix} 
 {\bf b}_{1,1}, \ldots, {\bf b}_{1,N} \\
 \cdots, \cdots, \cdots, \\ 
 {\bf b}_{N,1}, \ldots, {\bf b}_{N,N}.
\end{bmatrix},
\end{align}
where the $i$-th element of ${\bf b}_{k,j}$ is ${\bf b}_{k,j}^i=G_{{wd}_{ij}}G_{{wd}_{ki}}$.

Considering that the $d$-th term in the final expression without the numeric coefficient is denoted by $S_d({\bf x})$, the above expression can be expressed as
\begin{align}
    q_s({\bf x})=S_1({\bf x})+S_2({\bf x})+2S_3({\bf x}) + 4S_4({\bf x}) + S_5({\bf x}).
\end{align}
Considering $x_i^2=1$, the second moment of a quadratic form can be expressed as,
\begin{align}
    &{\rm E}[q_s({\bf x})]={\rm Tr}({\bf G})({\bf y}^T{\bf G}_{wd}{\bf y}+{\rm Tr}({\bf G})) + {\rm Tr}({\bf G}){\bf y}^T{\bf G}_{wd}{\bf y} +\notag \\ & 2{\rm Tr}({\bf Z}) + 4{\bf y}^T{\bf Z}_{wd}{\bf y} +\!\!\!\!\!\! \sum\limits_{l=1,i \neq j \neq k\neq l}^N\sum\limits_{k=1}^N\sum\limits_{j=1}^N\sum\limits_{i=1}^N y_i y_j y_k y_l G_{ij} G_{kl},
\end{align}
where ${\bf Z}={\bf G}_{wd}{\bf G}_{wd}^T$. This is readily found by taking expectation on \eqref{eq:secondm}. Note that the final term or $S_5({\bf y})$ can be found from the following observation:
\begin{align}
    S_5({\bf y})=q_s({\bf y})-\left(S_1({\bf y})+S_2({\bf y})+2S_3({\bf y}) + 4S_4({\bf y})\right).
\end{align}
The theorem is proved by combining the final two expressions.
\subsection{Proof of Corollary \ref{cor:grads}} \label{sec:gradsProof}
    As the gradient of most of the terms can be trivially calculated \cite{LaueMG2018,LaueMG2020}, we focus on the nontrivial gradient calculations here. In particular, the gradient of ${\bf y}^T{\bf U}_{wd}{\bf y}$ with respect to {\bf y} is derived next. We can write the following expression due to the chain rule:
\begin{align}
    \nabla_{\bf y} \left({\bf y}^T{\bf U}_{wd}{\bf y}\right)\! =&\! \begin{bmatrix}
        \sum\limits_k\sum\limits_j y_j y_k \frac{\partial}{\partial y_1} \left(({\bf U}_{wd})_{jk}\right) \\
        \sum\limits_k\sum\limits_j y_j y_k \frac{\partial}{\partial y_2} \left(({\bf U}_{wd})_{jk}\right) \\
        \vdots\\
        \sum\limits_k\sum\limits_j y_j y_k \frac{\partial}{\partial y_N} \left(({\bf U}_{wd})_{jk}\right)
    \end{bmatrix} +\notag\\&\! ({\bf U}_{wd}\!+\!{\bf U}_{wd}^T){\bf y},
\end{align}
where $({\bf U}_{wd})_{jk}$ is $(j,k)$-th element of the matrix ${\bf U}_{wd}$ and the matrix ${\bf U}$ can be expressed as
$${\bf U}=\begin{bmatrix} 
 ({\bf y}\odot{\bf y})^T{\bf b}_{1,1}, \ldots, ({\bf y}\odot{\bf y})^T{\bf b}_{1,N} \\
 \cdots, \cdots, \cdots, \\ 
 ({\bf y}\odot{\bf y})^T{\bf b}_{N,1}, \ldots, ({\bf y}\odot{\bf y})^T{\bf b}_{N,N}
\end{bmatrix}.  
$$
With this formulation, the inner derivative is $\frac{\partial}{\partial y_i} \left(({\bf U}_{wd})_{jk}\right) = 2y_i({\bf b}_{j,k})_i \quad \forall j\neq k$, where $({\bf b}_{j,k})_i$ is the $i$-th element of the vector 
${\bf b}_{j,k}$. Finally, the gradient can be written as,
\begin{align}
    &\nabla_{\bf y}\! \left({\bf y}^T{\bf U}_{wd}{\bf y}\right) = 2{\bf y}\! \odot\! \begin{bmatrix}
        \sum\limits_{k\neq j}\sum\limits_j y_j y_k ({\bf b}_{j,k})_1 \\
        \sum\limits_{k\neq j}\sum\limits_j y_j y_k ({\bf b}_{j,k})_2  \\
        \vdots\\
        \sum\limits_{k\neq j}\sum\limits_j y_j y_k \frac{\partial}{\partial y_N} ({\bf b}_{j,k})_N 
    \end{bmatrix} +\notag \\& ({\bf U}_{wd}+{\bf U}_{wd}^T){\bf y} \overset{(a)}{=} 2{\bf y} \odot {\bf b}_s +  ({\bf U}_{wd}+{\bf U}_{wd}^T){\bf y},
\end{align}
where $(a)$ is obtained by some matrix manipulations and the definition of ${\bf b}_{j,k}$ vectors.
\subsection{Proof of Lemma \ref{lem:bopt}} \label{sec:boptProof}
    We begin by calculating the total variance of the estimator $\tilde{\bf g}_{\bf p}$ below. %Define total variance operator
    \begin{align}
        &{\rm var}(\tilde{\bf g}_{\bf p})={\rm E}[\tilde{\bf g}_{\bf p}^T\tilde{\bf g}_{\bf p}] - {\rm E}[\tilde{\bf g}_{\bf p}]^T{\rm E}[\tilde{\bf g}_{\bf p}]\notag\\\overset{(a)}{=}&{\rm E}[\hat{\bf g}_{\bf p}^T\hat{\bf g}_{\bf p}]-{\rm E}[\tilde{\bf g}_{\bf p}]^T{\rm E}[\tilde{\bf g}_{\bf p}] - 2b_{\bf p}{\rm E}[\hat{\bf g}_{\bf p}^T {\bf d}_{\bf p}] + b_{\bf p}^2 {\rm E}[{\bf d}_{\bf p}^T{\bf d}_{\bf p}] \notag \\ 
        \overset{(b)}{=}& {\rm var}(\hat{\bf g}_{\bf p}) -  2b_{\bf p}{\rm E}[\hat{\bf g}_{\bf p}^T {\bf d}_{\bf p}] +   b_{\bf p}^2 {\rm var}({\bf d}_{\bf p}), \label{eq:vargp}
    \end{align}
where ${\rm var}({\bf d}_{\bf p})=   \frac{1}{N_e^2} \sum\limits_{j=1}^{N_e}{\rm var} \left(\nabla_{\bf p} \log \mathbb{P}_E( \boldsymbol{\theta}\{j\}|{\bf p},{\bf q})\right)$ and as the variance does not depend on the $j$-th index, we can calculate the variance of the inner quantity next ignoring the index.
\begin{align}
    &{\rm var} \left(\nabla_{\bf p} \log \mathbb{P}_E( \boldsymbol{\theta}|{\bf p},{\bf q})\right)\notag\\& = {\rm E}\left[ \left(\nabla_{\bf p} \log \mathbb{P}_E( \boldsymbol{\theta}|{\bf p},{\bf q})\right)^T \left(\nabla_{\bf p} \log \mathbb{P}_E( \boldsymbol{\theta}|{\bf p},{\bf q})\right)\right] \notag \\
    &={\rm E}\left[\sum\limits_{n=1}^N \left(\frac{\partial\log \mathbb{P}_E( \boldsymbol{\theta}|{\bf p},{\bf q})}{\partial p_n}\right)^2 \right] \notag\\&= \sum\limits_{n=1}^N {\rm E}\left[\left(\frac{\theta_n(\theta_n-1)}{2p_n}+\frac{(\theta_n^2-1)}{1-p_n-q_n}\right)^2\right] \notag \\
    &\overset{(a)}{=}  \sum\limits_{n=1}^N \left(\frac{1}{p_n} + \frac{1}{1-p_n-q_n}\right). \label{eq:vardp}
\end{align}
Using \eqref{eq:vardp} in \eqref{eq:vargp}, we can write that,
\begin{align}
    {\rm var}(\tilde{\bf g}_{\bf p})=& {\rm var}(\hat{\bf g}_{\bf p}) -  2b_{\bf p}{\rm E}[\hat{\bf g}_{\bf p}^T {\bf d}_{\bf p}] +\notag\\& \frac{ b_{\bf p}^2}{N_e}\sum\limits_{n=1}^N \left(\frac{1}{p_n} + \frac{1}{1-p_n-q_n}\right),
\end{align}
where $(a)$ is a result of the following observations: ${\rm E}[{\theta_n^{2k_0}}]=q_n+p_n$ and ${\rm E}[{\theta_n^{2k_0+1}}]=q_n-p_n$, where $k_0$ is a non-negative integer. Note that this is a convex quadratic expression in $b_{\bf p}$ and we can find the minimum by equating the derivative of this variance equal to zero. The optimal baseline is $b_{\bf p}^* = \frac{N_e}{\sum\limits_{n=1}^N \frac{1}{p_n} + \frac{1}{1-p_n-q_n}  } {\rm E}[ \hat{\bf g}_{\bf p}^T {\bf d}_{\bf p}]$ and the lemma is proved.

\hfill 
\bibliographystyle{IEEEtran}
\bibliography{hokie}

\end{document}